\documentclass[fleqn,usenatbib]{mnras}

\usepackage{newtxtext,newtxmath}

\usepackage[T1]{fontenc}

\DeclareRobustCommand{\VAN}[3]{#2}
\let\VANthebibliography\thebibliography
\def\thebibliography{\DeclareRobustCommand{\VAN}[3]{##3}\VANthebibliography}

\usepackage{graphicx}	
\usepackage{amsmath}	

\title[CZTI Compton spectroscopy]{Extending the energy range of {\em AstroSat}-CZTI up to 380 keV with Compton Spectroscopy}

\author[Kumar et al.]{Abhay Kumar$^{1,2}$\thanks{E-mail: abhaykk@prl.res.in},
Tanmoy Chattopadhyay$^{3}$,
Santosh V Vadawale$^{1}$,
A.R. Rao$^{4,5}$,
Mithun N. P. S.$^{1}$,
\newauthor
Varun Bhalerao$^{6}$,
and Dipankar Bhattacharya$^{4}$
\\
$^{1}$Physical Research Laboratory, Navrangpura, Ahmedabad, 380009, India.\\
$^{2}$Indian Institute of Technology, Gandhinagar, 382355, India.\\
$^{3}$Kavli Institute of Particle Astrophysics and Cosmology, 452 Lomita Mall, Stanford, CA 94305, USA.\\
$^{4}$The Inter-University Centre for Astronomy and Astrophysics, Pune, India.\\
$^{5}$Tata Institute of Fundamental Research, Mumbai, India.\\
$^{6}$Indian Institute of Technology Bombay, Mumbai, India.\\
}

\date{Accepted XXX. Received YYY; in original form ZZZ}

\pubyear{2021}

\begin{document}

\label{firstpage}
\pagerange{\pageref{firstpage}--\pageref{lastpage}}
\maketitle


\begin{abstract}

The CZTI (Cadmium Zinc Telluride Imager) onboard {\em AstroSat} is
 a high energy coded mask imager and spectrometer in the energy range of 20 - 100 keV. Above 100 keV, the dominance of Compton scattering cross-section in CZTI results in a significant number of 2-pixel Compton events and these have been successfully utilized for polarization analysis of Crab pulsar and nebula (and transients like Gamma-ray bursts) in 100 - 380 keV. These 2-pixel Compton events can also be used to extend the spectroscopic energy range of CZTI up to 380 keV for bright sources. However, unlike the spectroscopy in primary energy range, where simultaneous background measurement is available from masked pixels, Compton spectroscopy requires blank sky observation for background measurement. Background subtraction, in this case, is non-trivial because of the presence of both short-term and long-term temporal variations in the data, which depend on multiple factors like earth rotation and the effect of South Atlantic Anomaly (SAA) regions etc. We have developed a methodology of background selection and subtraction that takes into account for these effects. Here, we describe these background selection and subtraction techniques and validate them using spectroscopy of Crab in the extended energy range of 30 - 380 keV region, and  compare the obtained spectral parameters with the {\em INTEGRAL} results. This new capability allows for the extension of the energy range of {\em AstroSat} spectroscopy and will also enable the simultaneous spectro-polarimetric study of other bright sources like Cygnus X-1.

\end{abstract}




\section{Introduction}
The CZTI (Cadmium Zinc Telluride Imager) is a hard X-ray coded mask imager and spectrometer onboard India's first dedicated astronomy satellite \citep{singh14,kpsingh2022,paul13}, sensitive in the energy range of 20 - 100 keV. CZTI consists of four identical quadrants with 16 Cadmium Zinc Telluride (CZT) detector modules per quadrant. Each CZT module is 5 mm thick and 39.06 mm $\times$ 39.06 mm in dimension, totaling to a geometric area of 976 $cm^2$ \citep{vadawale16}. The CZT modules are pixelated in 16 $\times$ 16 array of pixels with a pixel size of 2.46 mm $\times$ 2.46 mm except for the edge row pixels having slightly smaller dimensions. CZTI is equipped with a 0.5 mm thick Tantalum coded aperture mask (CAM) for simultaneous imaging and background measurement. The CAM is supported by passive collimators and placed at a distance of 478 mm above the detector plane. Below each CZTI quadrant, there is a CsI(Tl) anti-coincidence detector (veto) at around 65 mm from the detector plane. The CsI crystal, with a thickness of 20 mm and cross-section of 167 mm $\times$ 167 mm, is read out by two Photo-Multiplier Tubes (PMT) from two sides of the crystal. The veto detector acts as the active shielding for the charged particle induced X-ray like signatures and the background gamma-ray radiation at higher energies (above 100 keV). More instrument details are given in \citet{bhalerao16}.

\par

CZTI data consists of an time-tagged event list with 20 $\mu$s timing resolution which include the information on the quadrant, CZT detector module ID, pixel number, and PHA value for each event. In addition flags to denote simultaneous detection in Veto detector is also available. CZTI data reduction pipeline takes this event list as an input and generate standard data products like light curves and spectra. For generation of background subtracted spectrum and light curves, CZTI analysis pipeline makes use of mask-weighting technique where the background is measured simultaneously considering the pixels open fraction and effective area (Mithun et al., in prep). This primary spectroscopic analysis is currently limited to the energy range of 30 - 100 keV because of the increasing mask transparency above 100 keV. With some modifications in the pipeline and proper accounting of the spatial variation of the background, it would, however, be possible to extend the primary spectroscopic energy range up to 150 keV. This work is under progress (Mithun et al., in prep). 

\par

Above 100 keV, because of relatively higher Compton scattering probability in the 5 mm thick CZT detectors, we see a sufficient number of 2-pixel Compton events thanks to the availability of event mode data with 20 $\mu$s resolution \citep{chattopadhyay14}.
The Compton events have been used for polarization measurement of Crab pulsar and nebula in 100 - 380 keV \citep{vadawale17} and a sample of Gamma-ray bursts \citep{chattopadhyay22grb,chattopadhyay19,chand18a,chand18b,Sharma_etal_2019,Sharma20,rahul2022}. Selection of the Compton events and polarization measurement have also been validated experimentally before the launch of {\em AstroSat} \citep{chattopadhyay14,vadawale15}. The same Compton events can be further used to extend the CZTI spectroscopy up to 380 keV.

	\par

	Sub-MeV Compton spectroscopy using the 2-pixel CZTI Compton events has been explored for GRBs in the past \citep{chattopadhyay21}. The availability of simultaneous background before and after the burst and significantly higher signal-to-noise ratio makes background subtraction and Compton spectroscopy for GRBs relatively easy. However, the same method can not be used in the case of on-axis sources like Crab. While mask weighting can be used for simultaneous background measurement in 30 - 100 keV, background subtraction using mask-weighting is not effective at higher energy because of the increasing transparency of the CAM above 100 keV. Therefore, for Compton spectroscopy of on-axis sources, separate background measurements are required using blank sky observations. X-ray background consists of multiple sources such as cosmic background radiation, induced radioactivity, trapped particles near the SAA, Earth Albedo, etc. \citep{riccardo2022}. A careful selection of blank sky observations is necessary to estimate the background due to all these sources correctly. The {\em AstroSat} satellite is in a near-Earth circular orbit of altitude 650 km and inclination of 6 degrees, thus making the cosmic ray induced background variation less drastic. The background, however, does vary due to multiple factors like the spacecraft's geometric location in orbit, the time spent within the South Atlantic Anomaly (SAA) region (a region with high intensities of trapped particles), and earth rotation. The background due to induced radioactivity will also depend on the time since the last SAA. Thus, the background depends on the ambient condition of the satellite outside the SAA and conditions within the past SAA crossings. The background variations due to SAA passages are short term variations. However, long term or quasi-diurnal variation is due to earth rotation \citep{antia2022}. These effects should be carefully incorporated while estimating the background. The low signal-to-ratio for on-axis sources makes background subtraction further complicated.

\par	
	
	We have developed a technique for Compton background selection and subtraction considering the short term and long term temporal variations in data. 
	Applying this technique, we have improved the spectroscopic sensitivity of the CZTI at higher energies for on-axis sources.

	In \citet{abhay2021}, we attempted sub-MeV spectroscopy (up to 700 keV) for Crab pulsar and nebula (hereafter referred to as `Crab') with the use of `low gain pixels' in CZTI. These are 20$\%$ of the total pixels with onboard post-launch gain found to be lower than the other CZTI pixels and hence are sensitive to photons up to 700 keV energy. With the use of these pixels, we attempted to extend energy range of the CZTI up to 700 keV. We could demonstrate that the CZTI does have sensitivity up to 700 keV. However, it was not possible to accurately constrain the spectral parameters, mainly because of the difficulty in estimating correct background. It was realized that while including the low-gain pixel does extend the energy range up to 700 keV, but the uncertainty of the background subtraction at the highest energies adversely affect the overall spectroscopic sensitivity. Therefore, in the present work we have decided to exclude the low-gain pixels and attempt the spectroscopy up to the energy of 380 keV, which is determined by the upper limit of the Compton event between two adjacent good pixels. Here we jointly analyse the Compton spectra obtained with the new background selection and subtraction methods along with the single event spectra to cover the wide energy range of 30 - 380 keV. We use all available clean Crab observations with CZTI having exposure greater than 30 ks, to establish the background selection and subtraction methods for the Compton spectroscopy.
	\par
	We compared our results with those obtained by the {\em INTEGRAL/SPI} satellite \citep{2009jourdain} and found them quite consistent with each other. This technique enables the spectroscopic study of  Crab like bright sources. There are many transient sources that reach such brightness levels; therefore, CZTI can be used for the extended spectroscopy of such sources. Since CZTI is sensitive to polarisation measurements, the simultaneous high energy spectro-polarimetric study of these bright sources can provide a wealth of information about the geometry and emission mechanism in these sources.
	\par
	This paper is structured as follows: in section \ref{sec_observation} and \ref{analysis}, observations and analysis procedure are briefly described, respectively. The results obtained are presented in section \ref{result}. Finally, in section \ref{discussion}, we discuss the Compton spectroscopy capability of CZTI and future plans. It is to be noted that this paper is primarily intended to describe the methods for Compton spectroscopy. The coded mask spectroscopy, which is available by default from the CZTI data pipeline, is the standard CZTI spectroscopy method and therefore has not been discussed in detail in this paper. However, some of the data cleaning procedures (described in sub-section \ref{short_term_variation}) are applied on both Compton and coded mask spectra.

\section{Observations}\label{sec_observation}

	\subsection{Crab observations}

\begin{figure}
		\centering
		\includegraphics[width=0.95\linewidth]{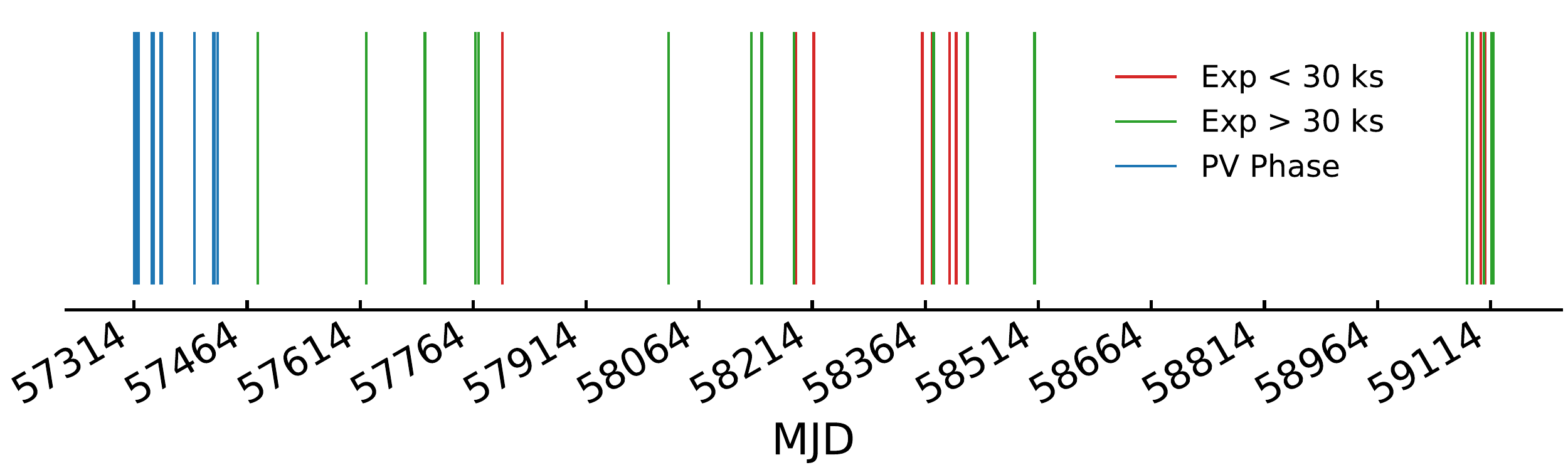}
		\caption{Crab observation times are plotted against MJD. Blue, red and green color represents PV phase observations, observations having exposure less than 30 ks and above 30 ks, respectively.} 
		\label{crab_obsid}
	\end{figure}

The Crab has been observed multiple times by {\em AstroSat} over the past six years, including in the Performance and Verification (PV) stage of CZTI. Figure \ref{crab_obsid} shows all the available Crab observations by CZTI. We have selected Crab observations 
having exposure greater than 30 ks to get a statistically significant number of Compton counts. 
	\begin{figure}
		\centering
		\includegraphics[width=0.95\linewidth]{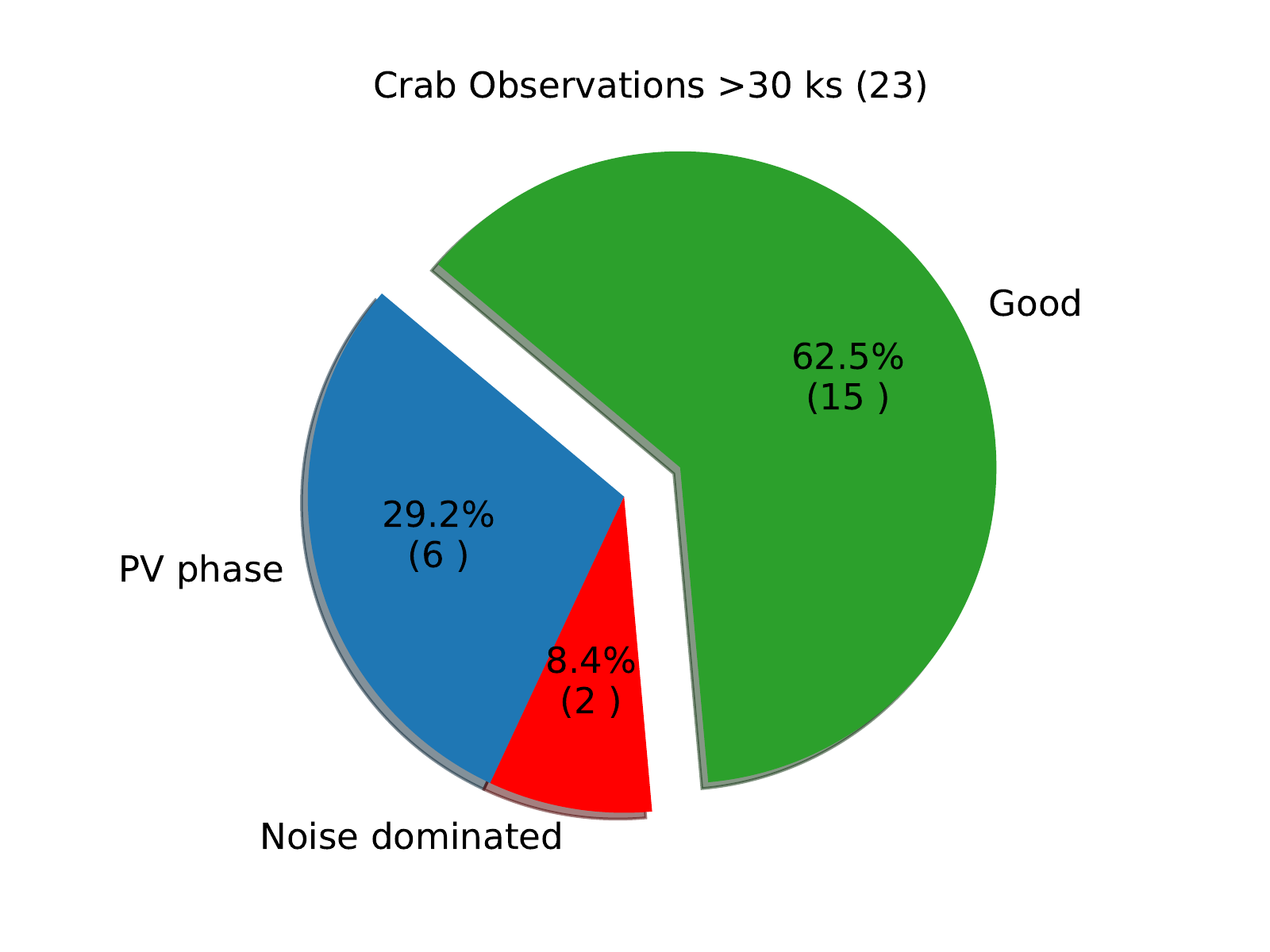}
		\caption{Crab observations distribution having exposure greater than 30 ks. 'Good' observations (15) are those after excluding the PV phase (6) and noise dominated observations (2). } 
		\label{pie_chart}
	\end{figure}
	For the present analysis, the observations with exposures below 30 ks (shown in red bars) are ignored because, during GTI (Good Time Interval) selection, almost half of the exposure is lost to filter out the charge particle background (more detail in section \ref{bkg_sel}) which results in an insufficient number of Compton events. Apart from this, we have excluded the PV phase observations (shown in blue bars) of the {\em AstroSat} i.e., September 2015 to March 2016 while the instrument configurations were being optimized. These considerations resulted in 17 observations between 2016 and 2020 for further analysis. Out of the 17 observations, two were found to be dominated by events from noisy pixels. When the number of events recorded by the instrument per second reaches its limit, subsequent events, including those from the source, are not recorded for the remaining fraction of the second. This happens on occasions when very noisy pixels are present making the data not very useful for scientific analysis. Therefore, we proceed with further analysis for the 15 observations (see Fig. \ref{pie_chart}). Details of the Crab observations (CZTI observation ID, date of observation, and time exposures) are given in Table \ref{observations}.

\begin{table*}
	\centering
	\caption{Summary of Crab and blank sky observations.}
	\label{observations}
	\begin{tabular}{lccccccr}
	
	\hline
	\multicolumn{3}{c}{Crab} & \multicolumn{5}{|c}{Blank Sky} \\
	\hline
		
	ObsID & Date & Exposure  & ObsID& Date & Exposure & RA & DEC \\
					&(yyyy/mm/dd) &(ks) & & (yyyy/mm/dd) & (ks) & (deg)& (deg)\\
					\hline
		
					$ 9000000406 $ & $ 2016/03/31 $ & $    114 $ & $ 9000004016 $ & $ 2020/11/16 $ & $     75 $ & $ 174.81 $ & $ 17.14 $ \\[0.2cm]
					
					$ 9000000620 $ & $ 2016/08/22 $ & $     84 $ &  $ 9000000956 $ & $ 2017/01/10 $ & $     81 $ & $ 183.48 $ & $ 22.80 $ \\[0.2cm]
					
					$ 9000000778 $ & $ 2016/11/ 8 $ & $     61 $ &  $ 9000000956 $ & $ 2017/01/10 $ & $     81 $ & $ 183.48 $ & $ 22.80 $ \\[0.2cm]
					
					$ 9000000964 $ & $ 2017/01/14 $ & $     78 $ &  $ 9000000956 $ & $ 2017/01/10 $ & $     81 $ & $ 183.48 $ & $ 22.80 $ \\[0.2cm]
					
					$ 9000000970 $ & $ 2017/01/18 $ & $    123 $ & $ 9000000956 $ & $ 2017/01/10 $ & $     81 $ & $ 183.48 $ & $ 22.80 $ \\[0.2cm]

					$ 9000001850 $ & $ 2018/01/15 $ & $    193 $
					& $ 9000002710 $ & $ 2019/02/07 $ & $     60 $ & $ 186.22 $ & $ 21.38 $ \\[0.2cm]
					
					$ 9000001876 $ & $ 2018/01/29 $ & $    234 $
					 & $ 9000002712 $ & $ 2019/02/08 $ & $     49 $ & $ 186.66 $ & $ 19.75 $ \\[0.2cm]

					$ 9000001976 $ & $ 2018/03/13 $ & $     41 $ & $ 9000002046 $ & $ 2018/04/22 $ & $     45 $ & $ 174.79 $ & $ 19.65 $ \\[0.2cm]
					
					$ 9000002368 $ & $ 2018/09/14 $ & $     49 $ &  $ 9000002712 $ & $ 2019/02/08 $ & $     49 $ & $ 186.66 $ & $ 19.75 $ \\[0.2cm]
					
					$ 9000002472 $ & $ 2018/10/29 $ & $     74 $ & $ 9000002712 $ & $ 2019/02/08 $ & $     49 $ & $ 186.66 $ & $ 19.75 $ \\[0.2cm]
					
					$ 9000002678 $ & $ 2019/01/26 $ & $    130 $
					&  $ 9000002710 $ & $ 2019/02/07 $ & $     60 $ & $ 186.22 $ & $ 21.38 $ \\[0.2cm]
					
					$ 9000003836 $ & $ 2020/08/22 $ & $     60 $ &  $ 9000000996 $ & $ 2017/01/31 $ & $     37 $ & $ 176.15 $ & $ 19.83 $ \\[0.2cm]
					
					$9000003848 $ & $ 2020/08/29 $ & $    290 $
					&  $ 9000002712 $ & $ 2019/02/08 $ & $     49 $ & $ 186.66 $ & $ 19.75 $ \\[0.2cm]
					
					$ 9000003900 $ & $ 2020/09/23 $ & $    136 $ &  $ 9000000996 $ & $ 2017/01/31 $ & $     37 $ & $ 176.15 $ & $ 19.83 $ \\[0.2cm]
					
					$ 9000003904 $ & $ 2020/09/26 $ & $    164 $ & $ 9000002710 $ & $ 2019/02/07 $ & $     60 $ & $ 186.22 $ & $ 21.38 $ \\[0.2cm]
					
					\hline
				\end{tabular}
	\end{table*}

\subsection{Background observations}

The selection of appropriate blank sky observation to measure the background for Compton spectroscopy is an essential part of the spectral analysis. The CZTI mask, collimators, and other support structures become increasingly transparent above 100 keV. Therefore, the actual blank sky background observation can be affected due to the presence of bright X-ray sources within 80 - 90 degrees of the pointing direction of CZTI. Special care should be taken, so that bright sources like Crab and Cygnus X-1 are outside the 80 - 90 degrees of the pointing direction during the background observation. The right ascension (RA) of Crab is 83 degrees, and Cygnus X-1 is 299 degrees which are almost 180 degrees away from each other. Therefore, it is possible to find a suitable region where both these sources are outside the 80 - 90 degrees region of the pointing direction. Apart from RA, we have also considered DEC (declination) of the background observation. We try to maintain the DEC within 5 degrees of the source observation to minimize any possible latitudinal variation of the background. Another important consideration is the time exposure of the blank sky observation which should be greater than 30 ks (same as source) to get a statistically significant number of Compton counts. Further, we have filtered out the observations which are having BAT sources with flux level $>$1.5 mCrab \citep{oh2018} in the field of view to avoid any possible contamination in the background.

	\begin{figure}
		\centering
		\includegraphics[width=0.95\linewidth]{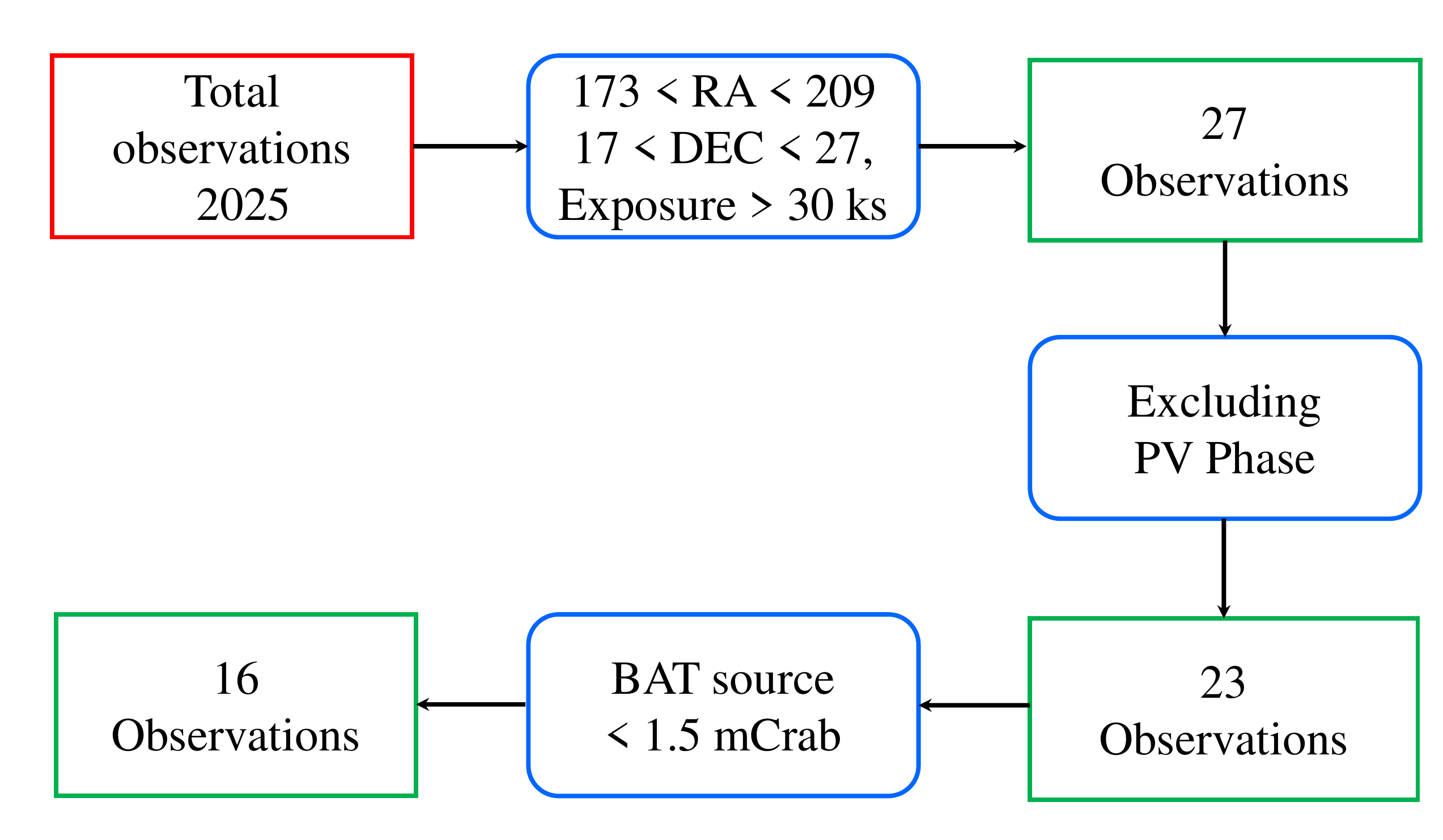}
	    \caption{Background selection flow chart. Total CZTI observations till 2020 is 2025 (shown in red). After applying the RA, DEC and exposure conditions, 27 are selected (shown in the 3rd box). Further, excluding PV phase observations resulted in 23 observations (shown in the 5th box). Finally, after applying the BAT flux criterion gives 16 blank sky observations as shown in the last box.} 
		\label{bkg_flow_chart}
	\end{figure}

After applying these criteria, we have selected 16 blank sky observations out of a total of 2025 observations to measure background. In Fig. \ref{bkg_flow_chart}, we show the different considerations behind the selection of sixteen blank sky observations in a flow chart. In Table \ref{observations}, we give the details of the selected blank sky observations, e.g., the CZTI observation ID, time exposures, and date of observation.

\section{Analysis procedure} \label{analysis}
 
\begin{figure}
		\centering
\includegraphics[width=1.05\linewidth,trim={0.3in 0.4in 0.5in 0.3in },clip]{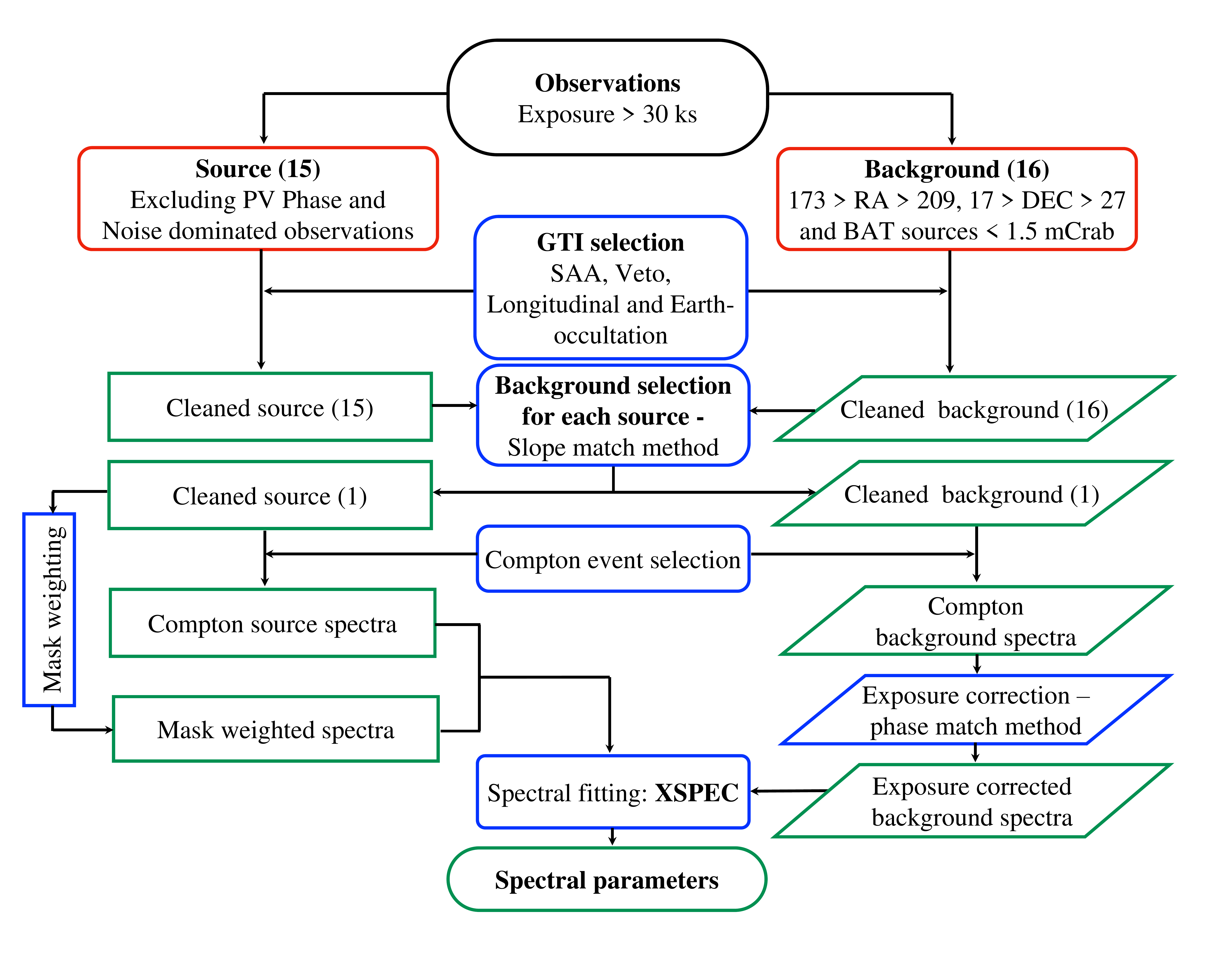}
		\caption{Crab spectral analysis flow chart. The red box shows the initial conditions for selecting 15 source regions and 16 background regions. The blue box is for the process, the green rectangular box represents the source outputs, and the green parallelograms represent the background outputs. } 
		\label{flow_chart}
	\end{figure}

 In this section, we discuss the methodology of Compton spectroscopy for Crab. The analysis steps are summarized in a flow chart in Fig. \ref{flow_chart}. We discuss these steps in the following subsections.

 \subsection{Selection of single-pixel and 2-pixel Compton events}{\label{event_sel}}

	 In this study, spectroscopic analysis of Crab is attempted in a broad energy range of 30 - 380 keV using two types of CZTI events $-$ the single-pixel events (hereafter `SE') in 30 - 100 keV and the 2-pixel Compton scattering events (hereafter `CS') in 100 - 380 keV. We remove events from noisy pixels (pixels having counts more than five sigma above the mean value) and spectroscopically bad pixels (poor energy resolution) to reduce the instrumental systematics. This is done for both the SE and CS events. The event selection criteria discussed below are applied to both the Crab and the blank sky background observations.

	 (1) Single-pixel events (SE):
	 
	A standard software, available at the AstroSat science support cell (ASSC) \footnote{\url{http://astrosat-ssc.iucaa.in/}}, is used to select the single-pixel events. These are isolated events in time;  there are no other events in a $\sim$40 $\mu$s time window around each of these events.

	(2) 2$-$pixel Compton events (CS): To identify the Compton events, first, we select the events occurring in a coincidence time window of 40 $\mu$s such that events must be recorded within 40 $\mu$s of each other in two separate pixels. The standard CZTI pipeline generates an event file with the list of the 2-pixel events.
	The actual Compton events (CS) are identified from the chance 2-pixel events by applying two other criteria: (1) spatial proximity of pixels $-$ events with only the adjacent pixels are considered, and (2) the sum and ratio of deposited energies must be consistent with the Compton kinematics. We estimate the polar scattering angle from the energies deposited in the two individual pixels. From the pixel geometry (2.5 mm pixel pitch and 5 mm in depth), we impose that the measured polar angle is within 26 to 154 degrees. Only those events with the sum of two energies between 100 and 380 keV are considered for analysis. The selection of Compton events has been discussed in the CZTI polarimetry papers in more detail $-$ \citet{chattopadhyay14,vadawale15,vadawale17,chattopadhyay19}. Compton event analysis is done outside the standard pipeline, and we have used algorithms written in Interactive Data Language (IDL).

\subsection{Data cleaning for short-term variation in data}\label{short_term_variation}
	
\begin{figure}
		\centering
		\includegraphics[width=1\columnwidth, trim={0. 0.1in 0.4in 0.35in },clip]{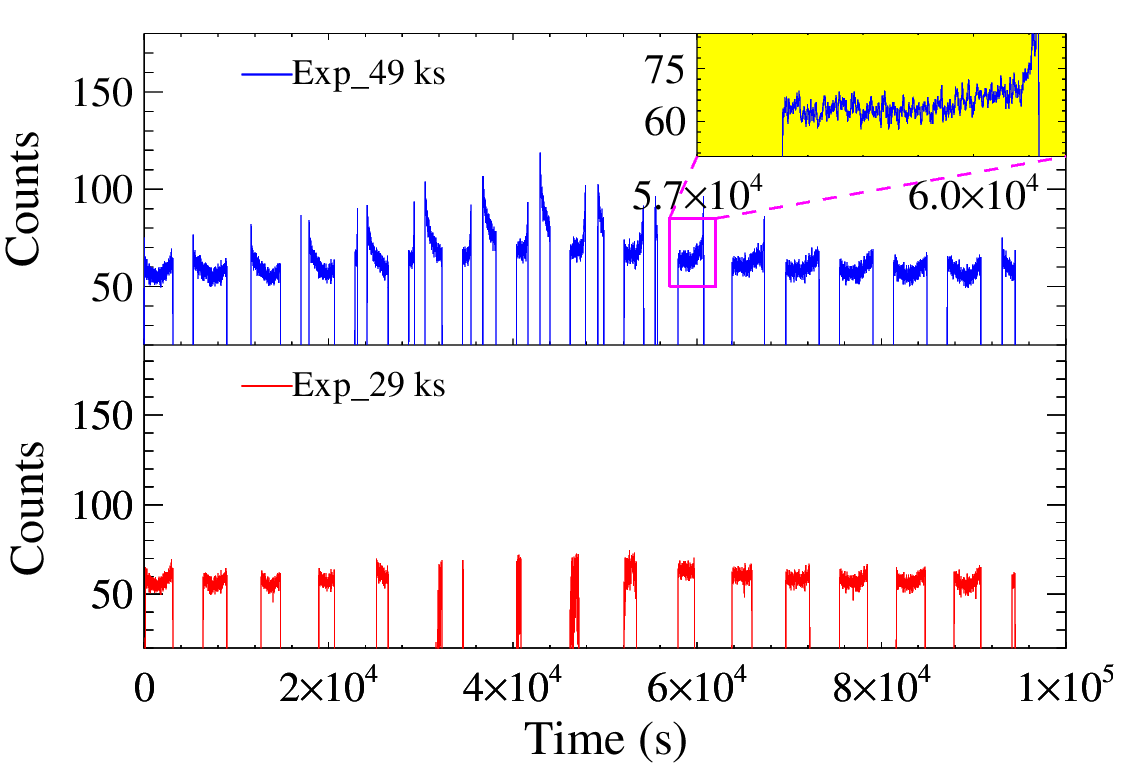}
		\caption{Light curve of the blank sky observation for Crab background (obsid: 2712). From top to bottom, the blue color represents raw light curve with SAA and earth-occultation cuts. The yellow box at the top right corner shows the zoom-in view of one of the orbits. The red light curve is after applying the longitudinal cuts and veto cuts.} 
		\label{lt_lc}
	\end{figure}

	Because of the interaction of charged particles with the spacecraft and the payload structure during the SAA passages, the background increases and decreases immediately upon the entrance and exit of the SAA \citep{riccardo2022}, which leads to short-term temporal variation in data. This can be seen in the individual orbital segments of the 1-s light curve (shown in the solid blue line in the top plot of Fig. \ref{lt_lc} and the inset plot) of Crab single-pixel events.
	Apart from the short-term variations, we also see a long-term quasi-diurnal variation in the data due to periodic variation in ground traces of the satellite because of earth rotation. More details of the long-term variations and their correction are discussed in subsection \ref{back_sub}. Here we describe the method to correct the short-term variation in the data.
 
	It is difficult to predict the particle background for an orbit in -135$^\circ$ to 45$^\circ$ degree of longitude because of multiple complex factors like the exact time span of satellite in the SAA regions, which is different for different orbits, depth in SAA regions, induced radioactivity in the detector, etc. Because of the unpredictable nature of the particle background, subtraction of blank sky from the source is impossible without correcting for this short-term variation in the data. Since the geomagnetic field in this longitudinal range (-135$^\circ$ to 45$^\circ$) is relatively weak, it leads to the high concentration of particle background \citep{ye17}. Thus, in our analysis, we have excluded the duration's from the GTI both for the Crab and blank sky background observations when the satellite is in the longitudinal range of -135 to 45 degree .  
	 
We also use counts in the CsI anti-coincidence (veto) detector as a proxy to identify high background regions. Using data from several long observations, we obtain histograms of veto count rates for each quadrant. It was seen that the distributions are asymmetric with a tail towards the higher count rate side corresponding to the observations near SAA or any other duration's of high background in orbit away from the SAA. We determine upper threshold for veto count rates of each quadrant such that they ignore typically 15$\%$ of the time corresponding to highest background levels while the background rates does not vary significantly during the remaining intervals. The thresholds for each quadrant are listed in Table \ref{vt_thrsld}. We ignore the time intervals where the counts in the veto detector of each quadrant are above these threshold values to remove the high particle background events anywhere in orbit away from the SAA.

	\begin{table}
	\centering
	\caption{Predefined Veto cut threshold values.}
	\label{vt_thrsld}
	\begin{tabular}{lccr} 
	\hline
	Quadrant & Threshold (counts-s$^{-1}$) \\
	\hline
	
	Q1 & 450\\
	Q2 & 500\\
	Q3 & 500\\
	Q4 & 480\\
	\hline
	\end{tabular}
	\end{table}
	
With these selection criteria, the regions of high charged particle background yielding the short-term variations in the light curve are removed. This is shown in the light curve in the solid red line in the bottom plot of Fig. \ref{lt_lc}. It is to be noted that we lose 40 - 50$\%$ of the original exposure in this process. We use the identical final GTI for data filtering in the single and 2-pixel Compton events for Crab and blank sky observations.

 \subsection{Selection of Background for Compton spectroscopy}\label{bkg_sel}

 In mask weighted single-pixel spectroscopy, the background is simultaneously measured during the source observation. However, because the mask weighting is effective up to 100 keV, measurement of Compton background (100 - 380 keV) requires a separate blank sky observation for each Crab observation. 
 
 \par
\begin{figure}
		\centering
		\includegraphics[width=0.95\linewidth]{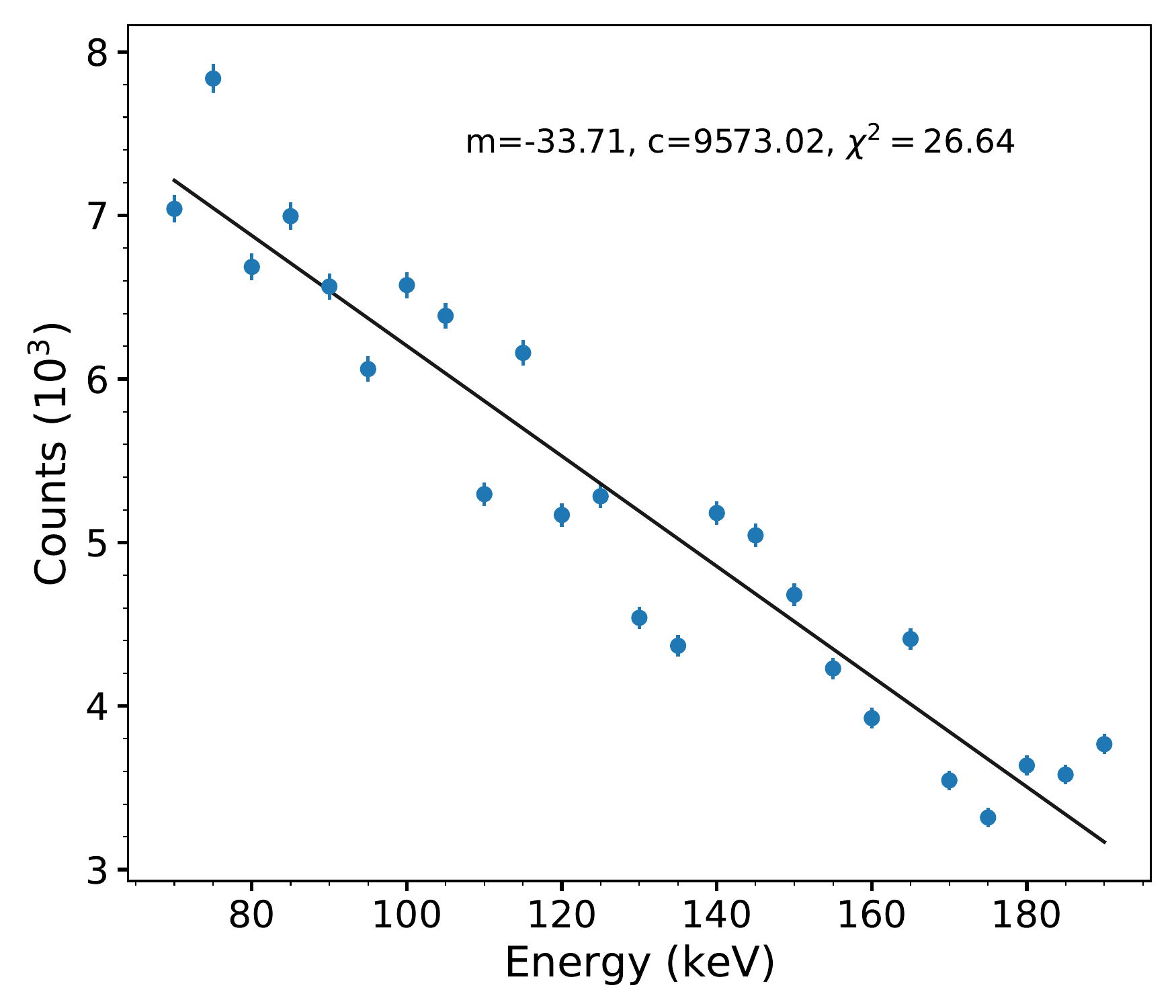}
		\caption{Spectral hardness using a straight line fit in 70 to 190 keV .} 
		\label{slope}
	\end{figure}

 Since a significant fraction of the observed counts in the source data consists of cosmic background, we can assume that variation of counts with the orbital position should be similar in the source and the blank sky observations. We also do not expect any variation in the source observation for Crab, which is known to be a steady source. To select the appropriate blank sky observation for each source observation, we compared the longitudinal variation of spectral hardness in background data with the spectral hardness in source data. The source and background observations are divided into six longitudinal bins starting from 45$^\circ$ to -135$^\circ$ longitude, with 30 degrees binning avoiding the SAA regions. The spectrum for each longitudinal bin is generated in 70 - 190 keV. The spectrum of each longitudinal bin is then fitted with a straight line where the fitted slope represents the hardness of the spectrum (see Fig. \ref{slope}). This is done for all the 15 Crab and 16 blank sky observations.  

\begin{figure}
		\centering
		\includegraphics[width=0.95\linewidth]{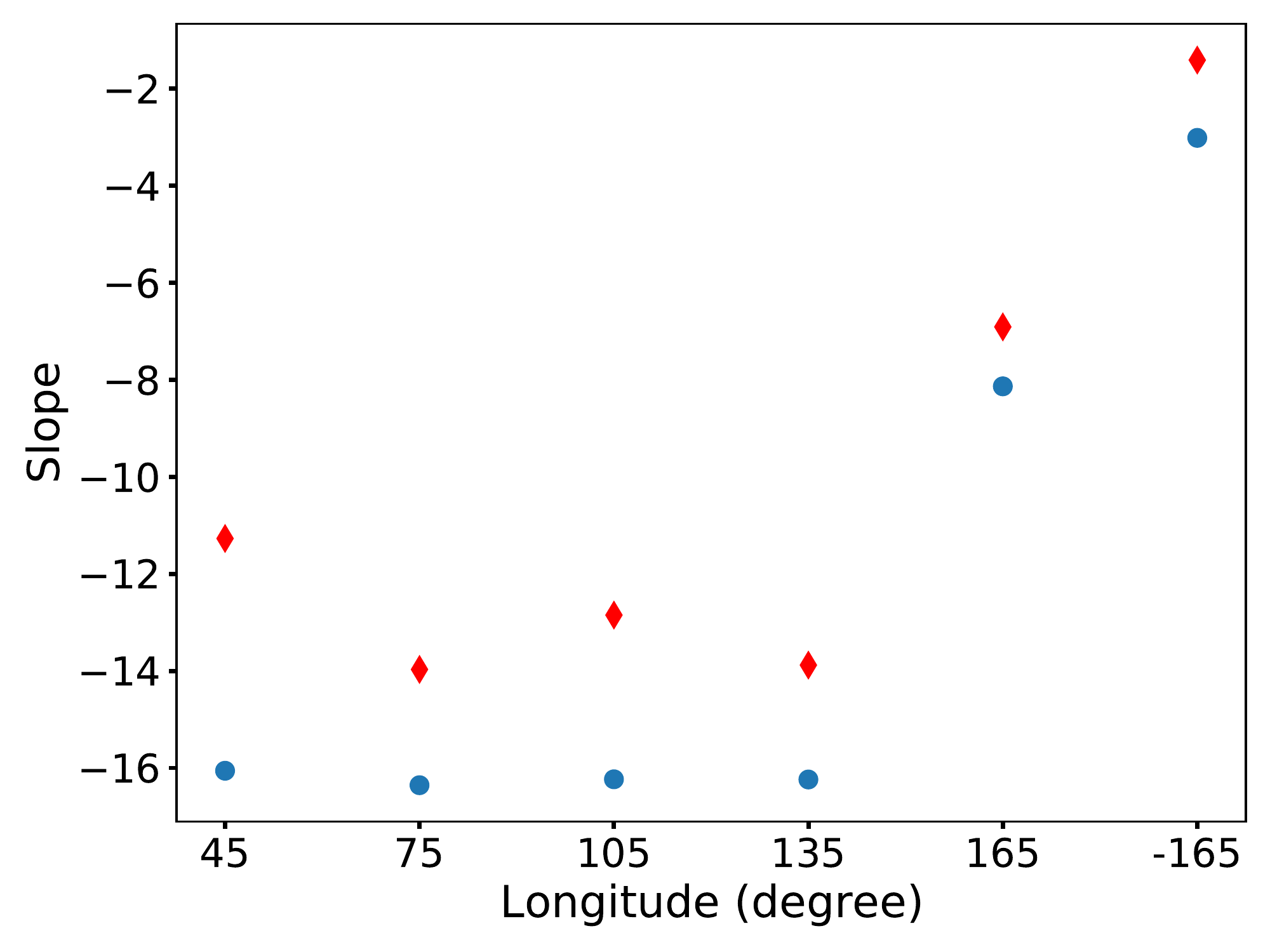}
		\caption{Longitudinal variation of the hardness of the Crab  and background. The blue filled circles represents Crab (obsid: 2472) and the red diamonds represents blank sky observation (obsid: 2712).} 
		\label{slop_var}
	\end{figure}
	\begin{figure}
		\centering
		\includegraphics[width=0.95\linewidth]{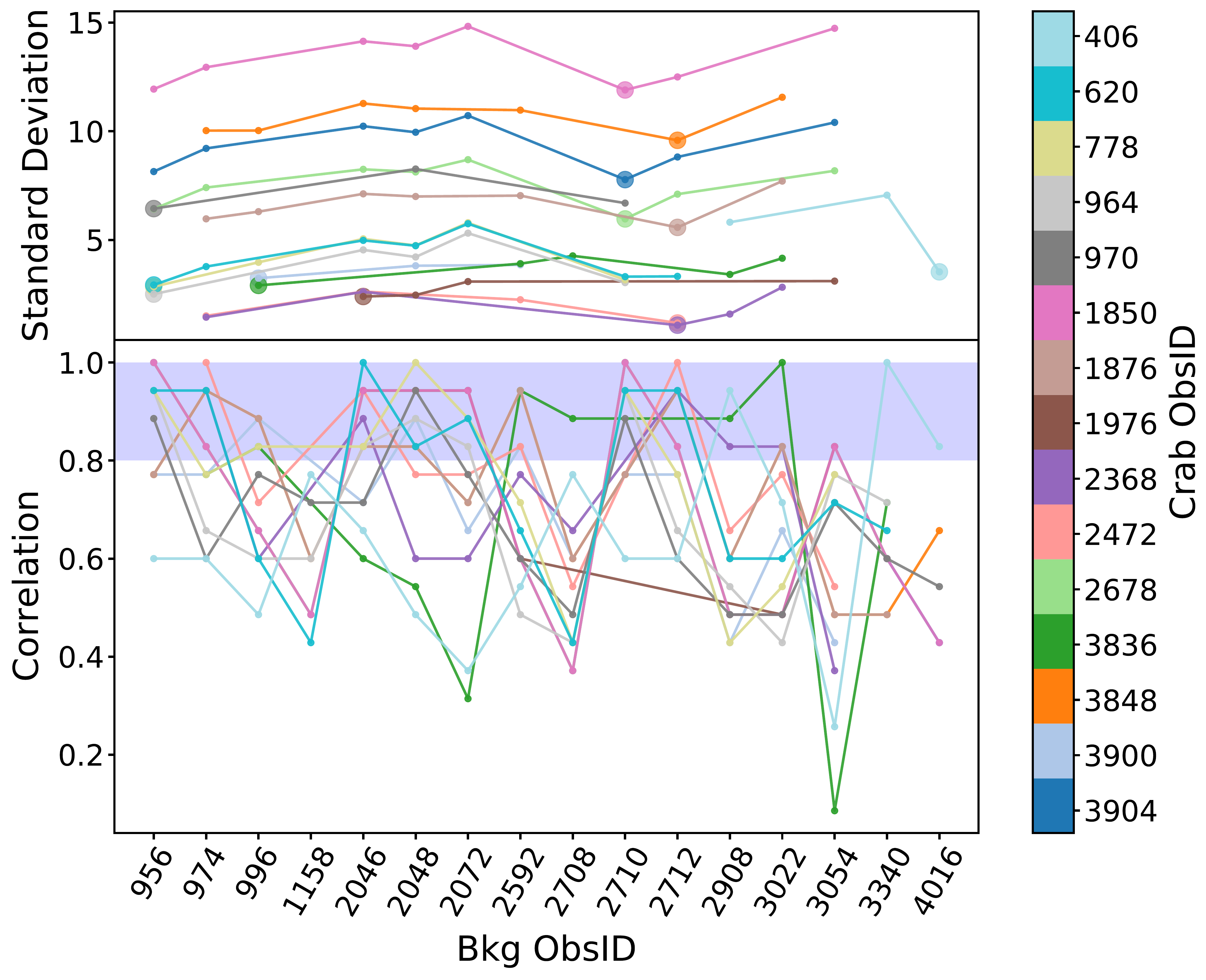}
        \caption{Bottom panel: correlation between the source and background longitudinal hardness variation for different Crab and blanks sky background observations. Top panel: standard deviation of the difference in the longitudinal hardness of the source and blank sky backgrounds having a correlation above 0.8. Each colour in the colour bar represents a different source observation. } 
		\label{bkg_cor}
	\end{figure}

In figure \ref{slop_var}, we show the longitudinal variation of the spectral hardness of the Crab (observation ID 2472) and blank sky observations (observation ID 2712). We expect the longitudinal variation of spectral hardness to be similar in the source and blank sky data for the correct blank sky for a given source observation. This is done in two steps: first, we calculated the correlation in longitudinal hardness variation between each source and all the background observations. All the backgrounds having a correlation greater than 0.8 for a given source observation are selected (see the bottom panel of Fig. \ref{bkg_cor}). We then calculate the standard deviation of the difference between the source and background slopes for the selected blank sky observations (see the upper panel of Fig.\ref{bkg_cor}). The variation of the difference should be minimum if the background embedded in the source is similar to the blank sky observation. Therefore, the blank sky observation for which the standard deviation is minimum is selected for background subtraction for that particular source observation. It is to be noted that at the beginning, we filter out the backgrounds for which the difference between the background and source slope in one or more longitudinal bins is negative. The final background is selected from the remaining background observations using the correlation and standard deviation algorithm.

\subsection{Exposure correction in Background due to long-term variation in data}\label{back_sub}

\begin{figure}
	\centering
	
	\includegraphics[width=0.95\linewidth]{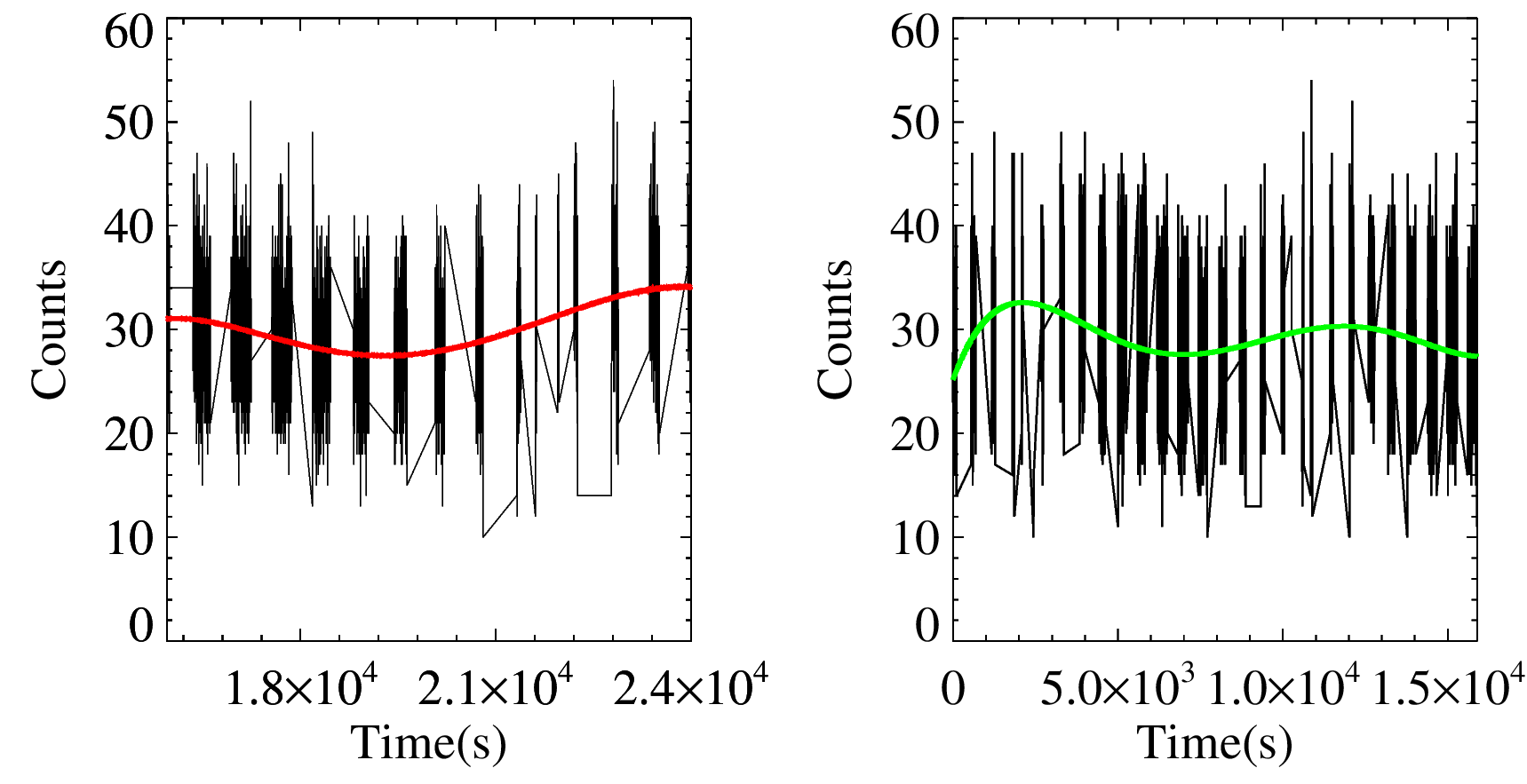}

	\includegraphics[width=0.95\linewidth]{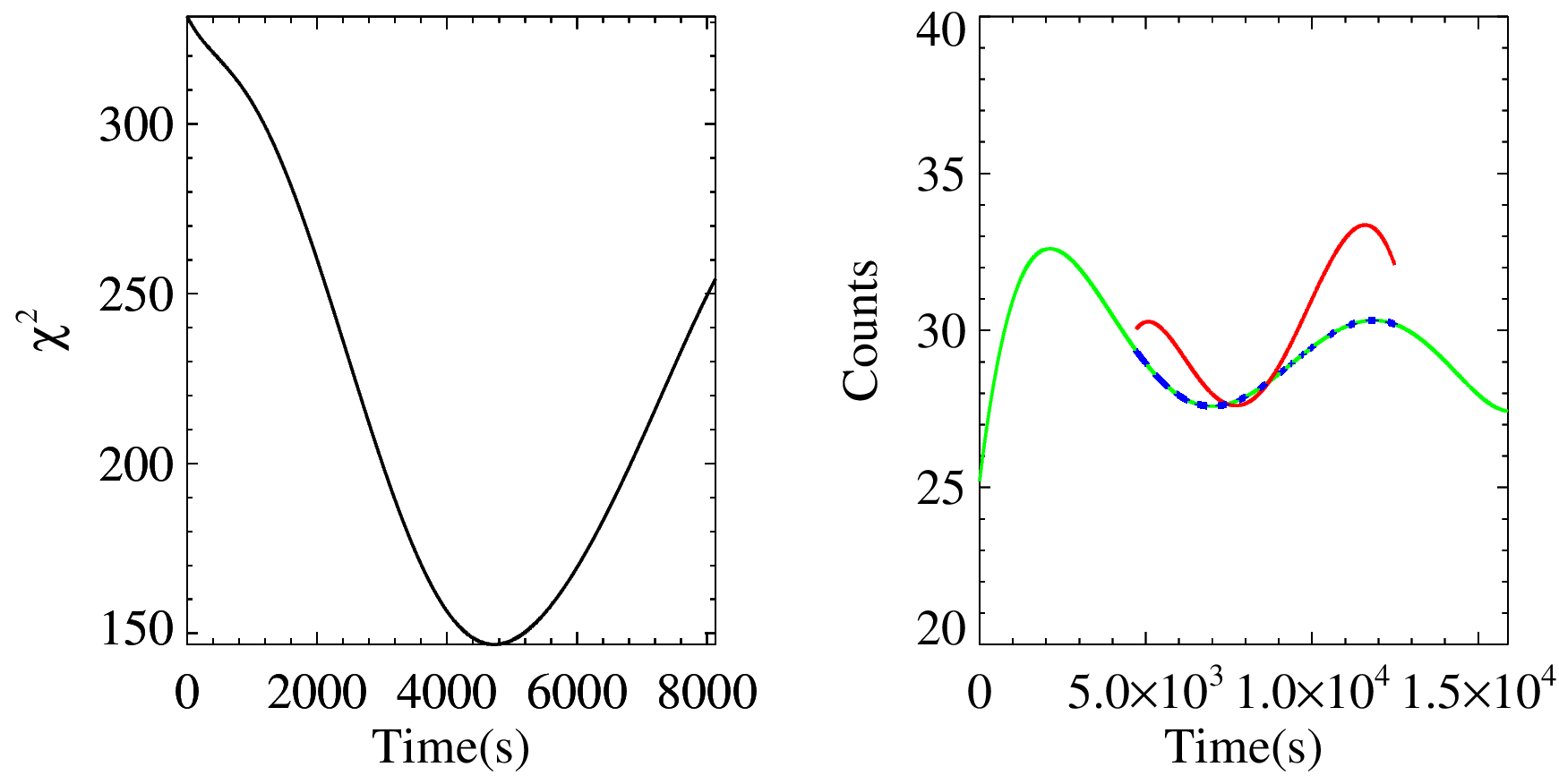}
	\caption{Top panel: light curves of Crab (a part of ObsId 970) on the left and background (ObsId 956) on the right, both fitted with fifth-degree polynomials (red curve for Crab and green curve for background). Bottom panel left figure shows the change of $\chi^{2}$ as a function of the shift in the Crab light curve with respect to the background to get the time for `phase matching'. The bottom panel right figure shows the `phase matched' background (duration of green curve overlapping with the red curve). The blue symbols over the phase matched background show the duration over which the Crab observations are available, and these are used to obtain the correction factor for the background. The gaps between blue symbols indicate the duration where there is no Crab data.} 
	\label{phasematch970}
\end{figure}

Here we describe methods to correct the long-term variation in the selected background data. An example of the long-term variation is shown in Fig. \ref{phasematch970} where we see a systematic modulation in the flux along the orbit of the satellite, arising because of the earth rotation \citep{antia2022}. Since the phases of such modulations in source and background are always different, subtraction of background without correcting for this will lead to incorrect net source flux. One way to address this issue is to obtain similar portions of the Crab and background orbits based on the spacecraft's ground tracks (latitude and longitude) for further analysis. However, this puts a stringent condition on orbital selection, resulting in a short usable exposure of Crab and background observations. An alternate method (the `phase match method') developed for background subtraction is to match the phase of the background and Crab light curves. The phase match method has been discussed and validated in detail in \citet{abhay2021}.

	In this method, first, we fit the source and background double event light curves in the 100 to 380 keV energy range with 5$^{th}$ degree polynomials. Then the source polynomial is shifted over the background polynomial every 10 seconds to compare and calculate the chi-square. The region with the minimum chi-square value is where the phases of the source and background match as shown in Fig. \ref{phasematch970}. The ratio of the total background count rate and the background count rate in the phase-matched region where there is Crab observation gives a correction factor to the background exposure, which is then used to scale the background count rate for subtraction. 
	The use of effective exposure, in this way, automatically takes care of the different phases of the source and background observations and thereby ensures that the source (Crab in this case) and blank sky background data are obtained for the same orbital phase (spacecraft orbit). For longer Crab observations (Crab exposure $>$ background exposure), the Crab light curve is divided into multiple parts and we calculate the correction factor for each of the parts as described above. The final correction factor is calculated as the weighted average of the individual correction factors.
	
\begin{figure}
		\centering
		\includegraphics[width=0.95\linewidth]{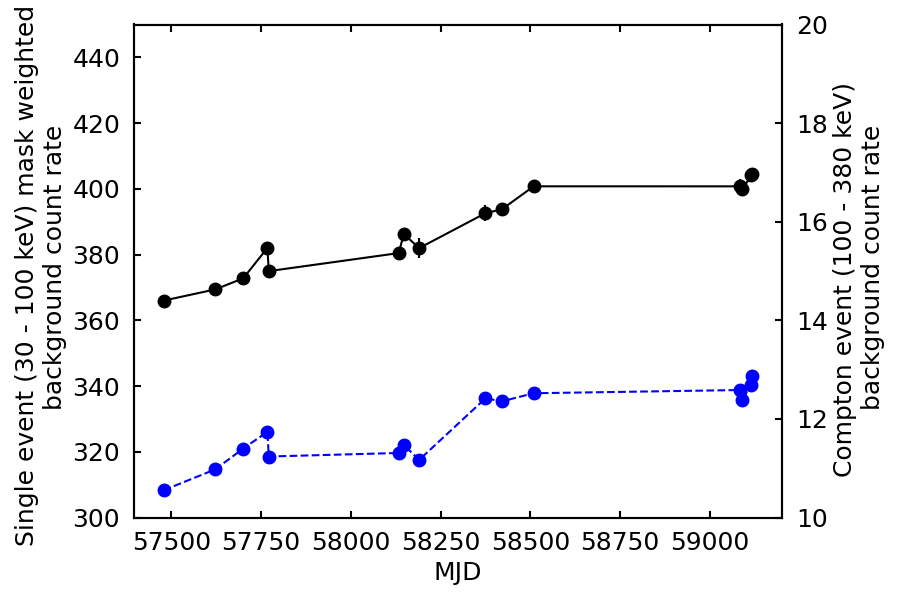}
		\caption{Mask weighted and Compton background count rate variation with MJD. The blue line with filled circles represents the Compton background (right hand side scale). The black  line with filled circles represents the mask weighted background (left hand side scale).} 
		\label{fig:compton_mask}
	\end{figure}

  In the previous subsections, we explained the selection of blank sky observation for the measurement of Compton background and methods to correct for the short-term and long-term variation in the data. To validate the measured Compton background (100-380 keV) and its variation with time, we compared it with the low energy background (30-100 keV) obtained by the mask weighting technique. We generated low energy background spectra (30 - 100 keV) for each observation by subtracting the source spectrum obtained by mask weighting from the total spectrum (source + background). Background count rates obtained from these spectra are shown with black data points in  Fig.\ref{fig:compton_mask}. 
The Compton background rates computed for each observation are shown in blue data points. Both the background rates are seen to be well correlated. The observed long-term secular variation in the background closely follows the trend of the number of Galactic Cosmic rays, anti-correlated with the Sunspot number that is decreasing towards its minimum in 2020/21 \citep{wang2022}. There are also additional short-term variations between observations that are also found to be correlated. The similar behavior of the estimated Compton event background and the low energy background strongly validates the background measurement method presented here.

\subsection{Generation of the Compton spectra}\label{response}
	
	Once the blank sky data for background measurement is selected for a given source observation and the background (and source) data are corrected for the long term variation, the next step is to generate the spectral and response files.
		\par

    For each of the observations, the single event spectrum (30 - 100 keV) is generated using {\tt cztbindata} module in the CZTI data analysis pipeline\footnote{\url{http://astrosat-ssc.iucaa.in/uploads/czti/CZTI\_level2\_software\_userguide\_V2.1.pdf}}. Clean event files with the GTI selection as described in previous sections are used as the input. For each event, the module assigns a mask-weight based on the open fraction of the respective pixel and the event energies are binned with respective weights to obtain background subtracted source spectrum (Mithun et al in prep). Spectra are generated with a bin size of 0.5 keV and corresponding response for spectral analysis is generated by using {\tt cztrspgen} module of the pipeline, which takes into account the spectral redistribution of CZT detectors based on $\mu \tau$ charge sharing model \citep{vadawale12, chattopadhyay16}. The mask-weighted spectroscopy with CZTI has been validated by simultaneous fitting of the Crab spectra from NuSTAR (see appendix \ref{SE_val}). The joint spectral analysis shows that the spectral index for the Crab spectra obtained with CZTI is consistent with that obtained form from NuSTAR. The present publicly available CZTI analysis pipeline shows a small difference in normalization of the two spectra, however, this has been accounted for with the better understanding of the systematics in the upcoming release of the CZTI pipeline (Mithun et al. 2022, in prep.)

	\par
	The 2-pixel Compton events are used to get the Compton scattering events spectrum in the 100 - 380 keV (CS) by estimating the sum of the energies deposited in the scattering and the absorption pixel. We have binned the data with a 10 keV bin size. The same steps are followed for the blank sky data to get background spectra in 100 - 380 keV. The GEANT4 simulation of the {\em AstroSat} mass model is used to generate the spectral response. Details of the mass model,  along with its validation, have been discussed in \citet{mate21} and \citet{chattopadhyay19}. In the simulation physics list, all the possible photon interactions with the detector are considered, such as photoelectric, Compton, Rayleigh, and scattering of photons from the spacecraft and the surrounding materials of the CZTI. First, the simulation of the mass model for 56 mono-energies starting from 100 keV up to 2 MeV  (at 20 keV interval up to 1 MeV and 200 keV till 2 MeV) is done.
\par

 It is done for a large number ($10^9$) of photons to get a statistically significant energy distribution for each mono-energetic beam. Selection of the 2-pixel Compton events is made using the same Compton kinematics criteria as discussed in section \ref{event_sel}. For the selected 2-pixel Compton events, the energies deposited in the corresponding two pixels are summed up to calculate the total deposited energy distribution with a bin size of 1 keV for each mono-energetic beam. We applied the CZTI pixel-level LLD (Lower Level Discriminator), the ULDs (Upper Level Discriminator), list of dead and noisy pixels obtained from the actual observational data to simulation data. We used a simple Gaussian model to convolve the measured energy distribution to calculate the Compton response. The Gaussian model has been considered in this case for simplicity in computation.

\section{Results} \label{result}

	We use an {\tt XSPEC} \citep{arnaud1996xspec} model, broken power-law, to fit the Crab curved spectra in 30 - 380 keV. It has long been used to explain Crab spectra with 100 keV break energy. \citep{strickman79,ling2003}. \citet{ling2003} reported using BATSE earth-occultation observations that the emission from the Crab in 35 - 300 keV can be described by a broken power-law with a break at ~100 keV. Later, {\em INTEGRAL/SPI} has also shown the spectral fitting using broken power law in the energy range of 23.5 keV - 1 MeV \citep{2009jourdain} with freezing break energy at 100 keV.
	In the case of CZTI, since the energy range is limited to 380 keV, we also considered freezing break energy at 100 keV for spectral fitting, as most of the instruments have considered.  
\par
	
\begin{figure}
		\centering
		\includegraphics[width=0.95\linewidth]{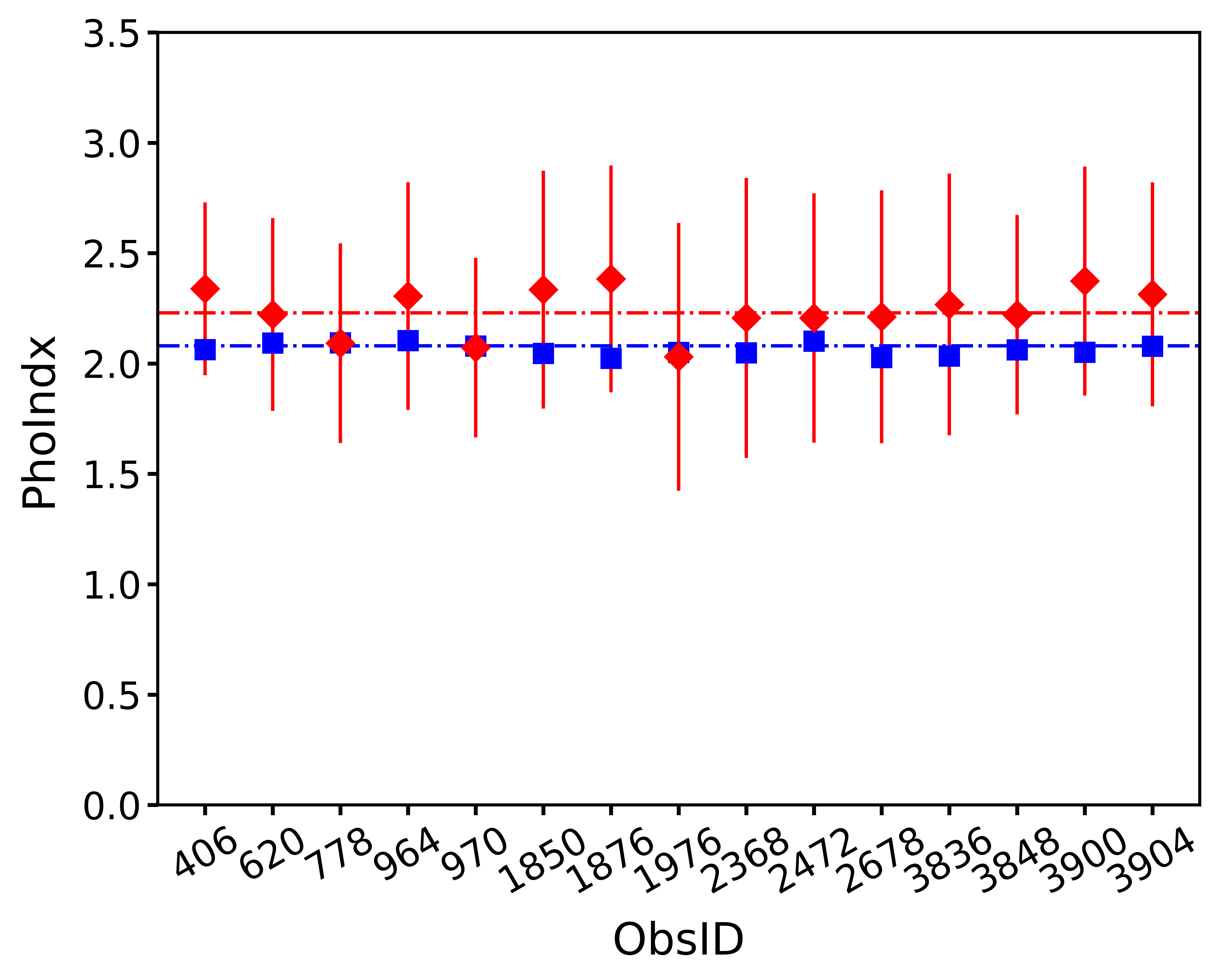}
		\caption{PhoIndx vs ObsId plot. The blue filled squares and red diamonds represents PhoIndx1 and PhoIndx2, respectively, and the dashed lines represents the {\em INTEGRAL/SPI} reported values.} 
		\label{parm}
	\end{figure}

\begin{figure}
		\centering
		\includegraphics[width=0.95\linewidth]{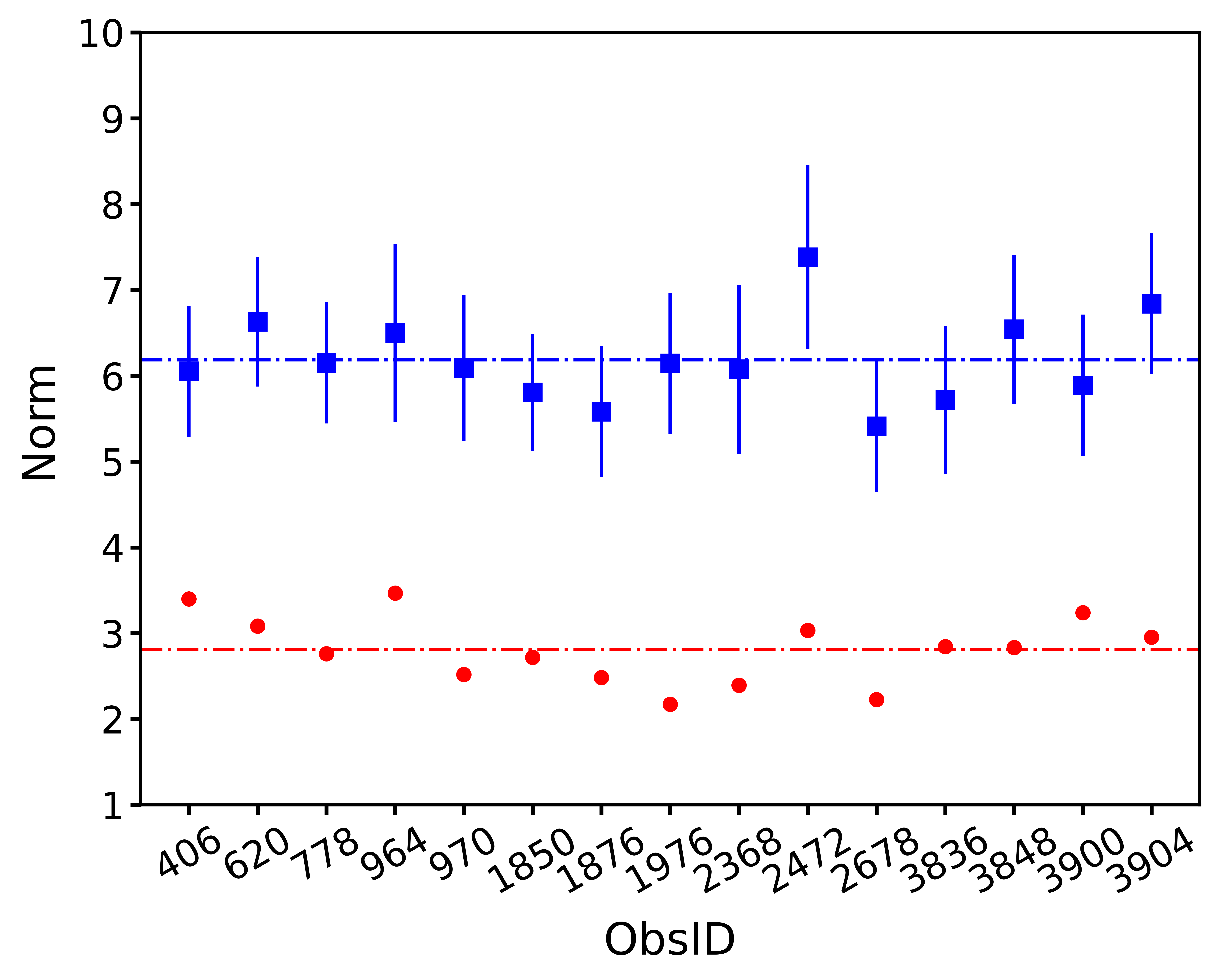}
		\caption{Norm vs ObsId. The blue filled squares and red filled circles represents SE and CS, respectively. The dashed lines represent the mean values.} 
		\label{norm}
	\end{figure}

	We fitted the CZTI SE and CS spectra simultaneously using const$\times$bknpower in {\tt XSPEC} freezing break energy at 100 keV (same as {\em INTEGRAL/SPI}). At the same time, the other parameters (photon indices) are tied across the spectra. A constant was multiplied to the model to account for the cross-calibration and differences between the different spectra. It was fixed to one for SE and left free to vary for CS. We have added a 4$\%$ systematic while fitting the observations having effective exposure $>$65 ks so that the reduced chi-square becomes less than 1.5. Spectral fitting for one of the 15 observations (ObsID 2368, 49 ks) is shown in Fig. \ref{2368spec}. The residual plots of the remaining observations are given in the appendix \ref{res_plot}. The fitted parameters for all the 15 observations along with the {\em INTEGRAL/SPI} results are given in Table \ref{integral_czti_results}. The low-energy slope ($PhoIndx1$) and the high energy slope ($PhoIndx2$) are well constrained and consistent with the {\em INTEGRAL/SPI} \citep{2009jourdain} results within error bars (see Fig. \ref{parm}).  The norm reported in Table \ref{integral_czti_results} correspond to SE. The value of the fitted normalisation constant for CS is consistently half compared to that for SE (see Fig. \ref{norm}),  which may be due to the inaccurate consideration of the effective area of the CZTI for the Compton response.

	\begin{figure}
		\centering
		\includegraphics[width=0.95\linewidth]{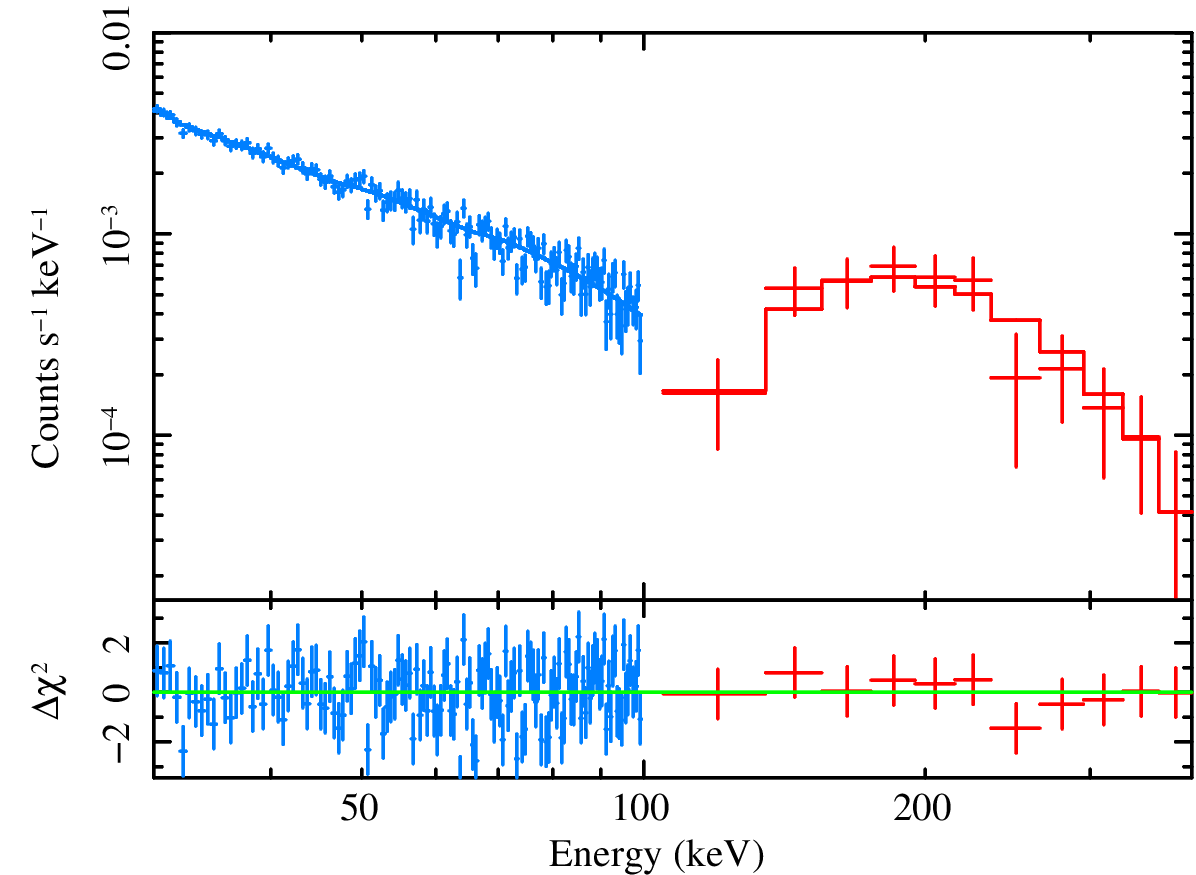}
	\caption{Broadband spectra of Crab (ObsID 2368, 49 ks) fitted with a broken power law. The blue and red colors are used for SE and CS, respectively.} 
		\label{2368spec}
	\end{figure}
	
\begin{table*}
	\centering
	
	\caption{Comparison of fitted parameters between {\em INTEGRAL/SPI} and {\em AstroSat} for broken power law. The errors are estimated for $90\%$ confidence interval. * represents a frozen parameter.}
				\label{integral_czti_results}

\begin{tabular}{lcccccccr}

\hline
Instrument&ObsID & Exposure & Effective &$PhoIndx1$ & $PhoIndx2$ &$E_{break}$& $Norm$  & $\chi^2$/dof\\
					& & &Exposure& & &   &(keV) &   \\
					\hline
				    NTEGRAL/SPI & & & $\sim$400 ks &$2.07^{+0.01}_{-0.01}$  &  $2.23^{+0.03}_{-0.03}$ &$100^{*}$ & &    \\[0.2cm]
				
			    	AstroSat/CZTI&$ 406 $ & 114 & 68 &$  2.06^{+0.03}_{-0.03} $ & $   2.34^{+0.40}_{-0.39} $ & $  {100*} $ & $    
				    6.05^{+0.89}_{-0.76} $ & $  188/163 $ \\[0.2cm] 
					&$ 620 $& 84& 53 & $  2.09^{+0.03}_{-0.03} $ & $   2.22^{+0.44}_{-0.44} $ & $  {100*} $ & $    6.63^{+0.87}_{-0.75} $ & $  209/163 $ \\[0.2cm] 
					&$ 778 $ & 61& 37 & $  2.09^{+0.03}_{-0.03} $ & $   2.09^{+0.48}_{-0.45} $ & $  {100*} $ & $    6.15^{+0.81}_{-0.71} $ & $  228/161 $ \\[0.2cm] 
					&$ 964 $ & 78& 44& $  2.10^{+0.05}_{-0.04} $ & $   2.31^{+0.54}_{-0.52} $ & $  {100*} $ & $    6.50^{+1.26}_{-1.04} $ & $  205/163 $ \\[0.2cm] 
					&$ 970 $ & 123 & 72& $  2.08^{+0.04}_{-0.04} $ & $   2.07^{+0.41}_{-0.41} $ & $  {100*} $ & $    6.09^{+1.00}_{-0.85} $ & $  184/163 $ \\[0.2cm] 
				    &$ 1850 $ &193& 105& $  2.05^{+0.03}_{-0.03} $ & $   
				    2.33^{+0.56}_{-0.54} $ & $  {100*} $ & $    
				    5.81^{+0.81}_{-0.68} $ & $  232/163 $ \\[0.2cm] 
				    &$ 1876 $ &234& 112& $  2.03^{+0.04}_{-0.04} $ & $   2.38^{+0.52}_{-0.51} $ & $  {100*} $ & $    5.58^{+0.93}_{-0.77} $ & $  243/163 $ \\[0.2cm]
					&$ 1976 $ &41& 24& $  2.05^{+0.04}_{-0.04} $ & $   2.04^{+0.62}_{-0.61} $ & $  {100*} $ & $    6.14^{+1.01}_{-0.82} $ & $  205/163 $ \\[0.2cm] 
					&$ 2368 $ &49&28& $  2.05^{+0.05}_{-0.05} $ & $   2.21^{+0.65}_{-0.63} $ & $  {100*} $ & $    6.08^{+1.20}_{-0.98} $ & $  207/163 $ \\[0.2cm]
					&$ 2472 $ &74&33& $  2.10^{+0.04}_{-0.04} $ & $   2.21^{+0.60}_{-0.56} $ & $  {100*} $ & $    7.38^{+1.28}_{-1.07} $ & $  203/163 $ \\[0.2cm]
					&$ 2678 $ &130&34& $  2.03^{+0.04}_{-0.04} $ & $   2.21^{+0.60}_{-0.57} $ & $  {100*} $ & $    5.41^{+0.92}_{-0.77} $ & $  230/161 $ \\[0.2cm]
					&$ 3836 $ &60&29& $  2.03^{+0.04}_{-0.04} $ & $   2.27^{+0.61}_{-0.59} $ & $  {100*} $ & $    5.72^{+1.05}_{-0.87} $ & $  202/161 $ \\[0.2cm]
					&$ 3848 $ &290&157& $  2.06^{+0.04}_{-0.04} $ & $   2.22^{+0.47}_{-0.45} $ & $  {100*} $ & $    6.54^{+1.04}_{-0.87} $ & $  216/161 $ \\[0.2cm]
					&$ 3900 $ &136&40& $  2.05^{+0.04}_{-0.04} $ & $   2.37^{+0.54}_{-0.52} $ & $  {100*} $ & $    5.89^{+1.00}_{-0.82} $ & $  205/161 $ \\[0.2cm] 
					&$ 3904 $ &164&64& $  2.08^{+0.03}_{-0.03} $ & $   2.31^{+0.54}_{-0.51} $ & $  {100*} $ & $    6.84^{+0.95}_{-0.82} $ & $  221/161 $ \\[0.2cm]

				\hline

	\end{tabular}
	
\end{table*}

\section{Summary} \label{discussion}

In this article, we have outlined methods for Compton spectroscopy using \emph{AstroSat}-CZTI. We use the single-pixel mask-weighted spectral data in the 30 - 100 keV energy range (SE) and 2-pixel Compton events in the 100 - 380 keV energy range (CS) for a joint 30 - 380 keV broadband spectroscopy. As a proof of concept, we applied this analysis for Crab and compared the spectral parameters with the {\em INTEGRAL/SPI} reports which were found to be consistent. 
\par
In the case of on-axis sources, the unavailability of simultaneous background measurement for Compton spectroscopy makes the estimation of accurate blank sky flux and its subtraction essential, mainly when the signal-to-noise ratio is relatively low. The selection of blank sky data is done by comparing the longitudinal variation of the spectral hardness of the source and blank sky data (subsection \ref{bkg_sel}). However, the presence of short-term and long-term temporal variation in the observed count rates due to high charge particle concentration in the SAA regions and earth rotation, respectively, make the subtraction of background non-trivial and complicated. To exclude the short-term variation, we ignore the data from  -135$^\circ$ to 45$^\circ$ longitude, where the earth's magnetic field is relatively weak compared to other regions (subsection \ref{short_term_variation}). The long-term variation, on the other hand, is corrected by comparing the phases of the Crab and the selected background and scaling the overall background count rate according to the count rate observed in the matched phase (subsection \ref{back_sub}). While this method ensures correct estimation of background (and source) flux, the background selection procedure makes sure that the selected blank sky background and the original background embedded in the source observation follow the same spectral energy distribution.
\par
For this work, the Compton spectral response was generated by simulating the {\em AstroSat} mass model in Geant4. The estimated energy distributions were convolved with a Gaussian kernel for simplicity (section \ref{response}), which we intend to improve with the use of a physical line profile model based on charge trapping and diffusion \citep{chattopadhyay16}. In the future, we would also include the charge sharing effect in Compton response for improved spectral analysis. For CZTI like pixelated detectors, charge sharing between the adjacent pixels is expected to give rise to 2-pixel events that mimic the Compton events. This effect has been seen in the polarimetry data analysis of GRBs with CZTI (Chattopadhyay et al., 2022, under review) as well as in polarimetry experiments with CZT detectors identical to those used in CZTI (Vaishnav et al., 2022, under review). Inclusion of charge sharing in the response will help in better estimation of the relative norm (and source flux in 100 - 380 keV) for the Compton spectra. 
\par
We have done the spectral analysis of the Crab using these techniques, where we used a broken power-law (bknpower in {\tt XSPEC}) for the spectral fitting. The spectral fits showed that CZTI has sufficient flux sensitivity to perform spectroscopy for on-axis bright sources up to 380 keV (see Table \ref{integral_czti_results}). For the single event and Compton event spectra in 30 - 380 keV (SE and CS), the low-energy slope ($PhoIndx1$) and higher energy slope ($PhoIndx2$) agree reasonably with the {\em INTEGRAL/SPI} result.

\par
To summarize, with the background selection and subtraction methods, described in this paper, we find that CZTI has the capability to do spectroscopy for on-axis sources in 30 - 380 keV. This will provide sensitive spectroscopic information for various bright ($\sim$1 Crab) hard X-ray sources such as Cygnus X-1 and other transient sources, which will help in better understanding of the emission mechanisms in these sources. CZTI is also a sensitive polarimeter; hence, simultaneous spectroscopy and polarisation measurements can give a wealth of information about these sources. For example, for bright black hole binaries like Cygnus X-1, a detailed spectro-polarimetry study will help us understand the putative jet's emission mechanism, geometry, and magnetic field structure.

\section*{Acknowledgements}

This research is supported by the Physical Research Laboratory, Ahmedabad, Department of Space, Government of India. We acknowledge the ISRO Science Data Archive for {\em AstroSat} Mission, Indian Space Science Data Centre (ISSDC) located at Bylalu for providing the required data for this publication, and Payload Operation Center (POC) for CZTI located at Inter-University Centre for Astronomy $\&$ Astrophysics (IUCAA) at Pune for providing the data reduction software.


\section*{Data Availability}
We have used data available at the ISRO Science Data Archive for {\em AstroSat} Mission, Indian Space Science Data Centre (ISSDC) located at Bylalu.




\bibliographystyle{mnras}
\bibliography{reference} 




\appendix

\section{Validation of Coded mask spectroscopy}\label{SE_val}
 
\begin{figure}
		\centering
\includegraphics[width=0.85\linewidth,trim={0. 0.2cm 0. 0.}]{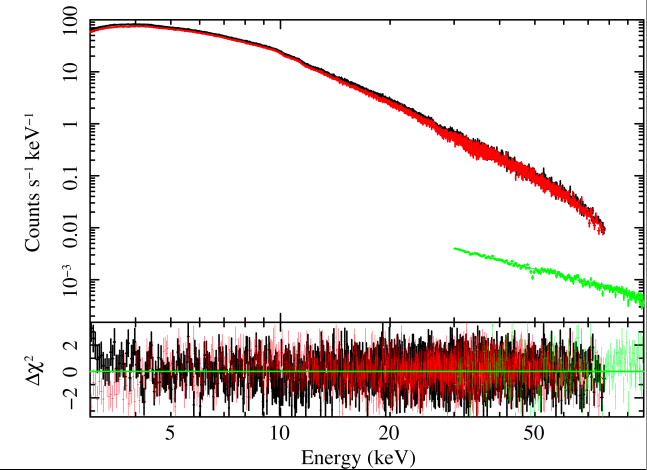}
\includegraphics[width=0.85\linewidth,trim={0. 0. 0. 0.}]{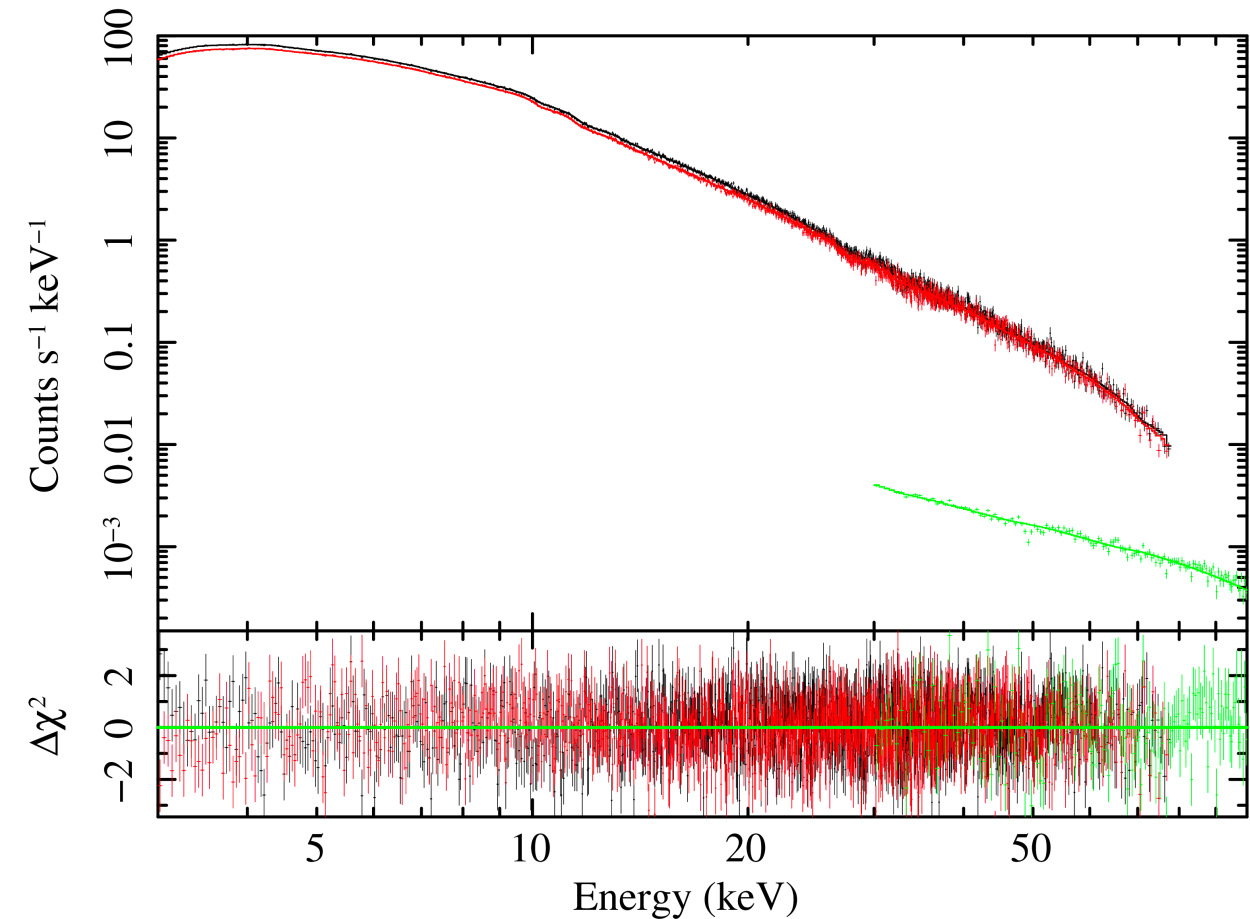}

\caption{Nustar and SE simultaneous fit (Top: Case 1, Bottom: Case 2). Black is FPMA, red is FPMB, and green is the SE spectrum. } 
\label{nu_val}
\end{figure}

 We have one Crab observation (ObsID 406) for which there is strictly simultaneous NuSTAR observation (nu10002001009) available. So, we used this observation for validation of SE by joint fitting. The SE is obtained using the existing CZTI data reduction pipeline. We have carried out simultaneous spectral fitting of the SE and the NuSTAR (FPMA and FPMB) Crab spectrum (3-79 keV) (see Fig. \ref{nu_val}) in XSPEC using tbabs $\times$ pow model  where $N_{H}$ is frozen to $2.2\times10^{21}$ cm$^-{2}$ \citep{madsen2017}. The spectral fitting is done in two different ways. Case 1: we have tied all the parameters, and to account for cross-calibration differences, a normalisation constant is multiplied to the model. It is frozen to one for FPMA and left free for the others instruments. Case 2: we have frozen the normalisation constant obtained from Case 1, and all other parameters are left free (see Table \ref{czti_nu_val}). In all the cases, we get a photon index consistent with the results from NuSTAR \citet{madsen2017}. The norm for CZTI is found to be slightly lower than that of NuSTAR. The existing norm discrepancies will be taken care of in the new CZTI pipeline with better handling of the systematics and background. The new results and the details of coded mask, coded mask spectroscopy method and response generation are given in Mithun et al.,(under prep).

\begin{table*}
\centering
\caption{Comparison of fitted parameters between {\em NuSTAR} and {\em AstroSat} for power law. The errors are reported for a $90\%$ confidence interval. * represents a frozen parameter.}\label{czti_nu_val}
	
\begin{tabular}{lcccccccr}
\hline

Pow paramters & Cases & Nustar (FPMA)& Nustar (FPMB)&CZTI & $\chi^2$/dof & \\
 
 \hline 
 PhoIndex& tied &2.108$\pm$0.001 & & & 2714/2381& \\[0.2cm]
                     Norm & tied &9.40$\pm$0.03 & & & & \\[0.2cm]
                     Constant &free & $1^{*}$& $0.97$&$0.79$ & \\[0.2cm]
 
  \hline
  PhoIndex& free & 2.113$\pm$0.002 &2.103$\pm$0.002 & 2.091$\pm$0.021 & 2680/2379 & \\[0.2cm]
                 Norm &free & 9.49$\pm$0.04& 9.31$\pm$0.04 &$8.80^{+0.76}_{-0.69}$ & \\[0.2cm]
         Constant &frozen to above value & $1^{*}$ &$0.97^{*}$ &$0.79^{*}$ & \\[0.2cm]
\hline

\end{tabular}
\end{table*}

\section{Residual Plots}\label{res_plot}

\begin{figure}
		\centering
		\includegraphics[width=0.95\linewidth]{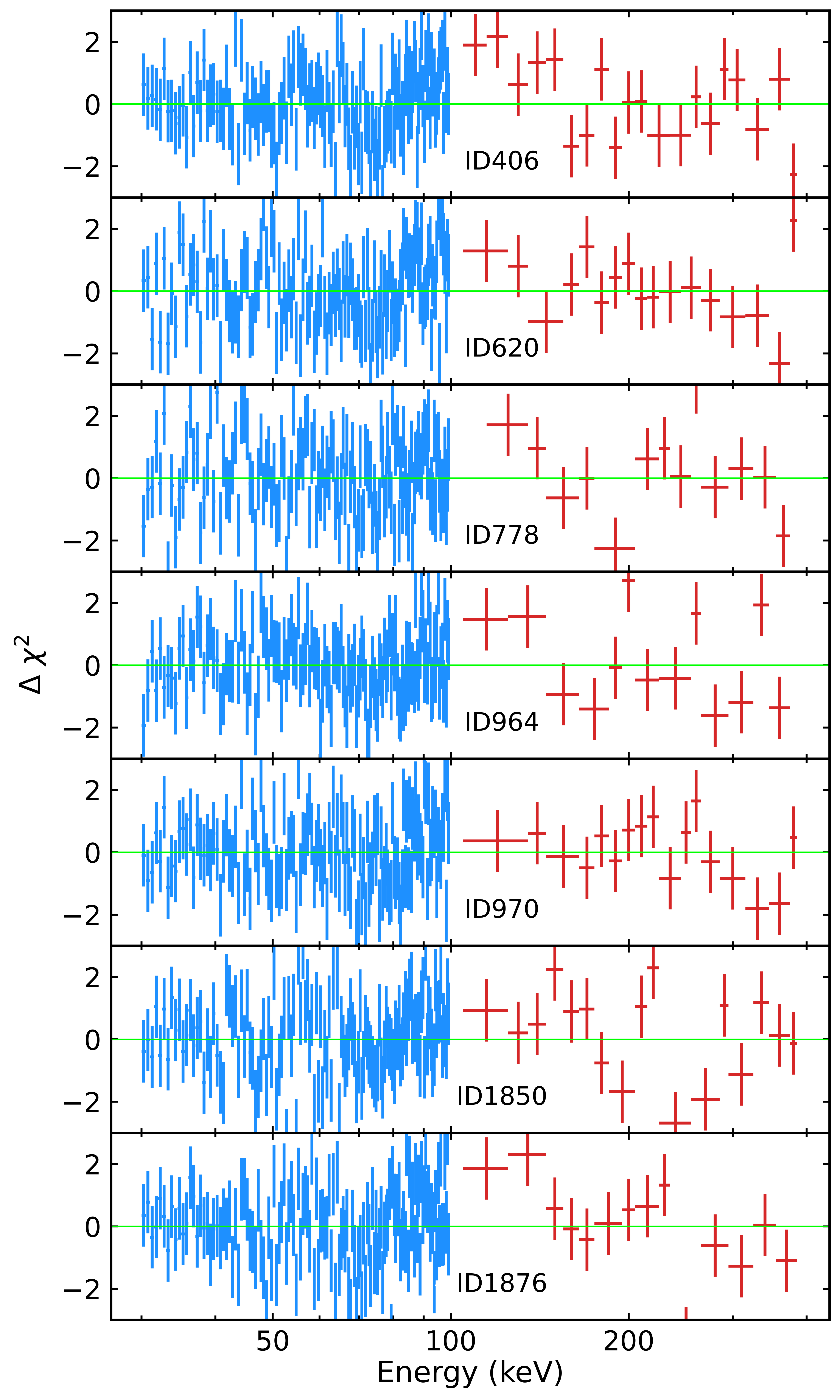}
		\end{figure}
		
		\begin{figure}
		\centering
\includegraphics[width=0.95\linewidth,trim={0. 0. 0. 0.}]{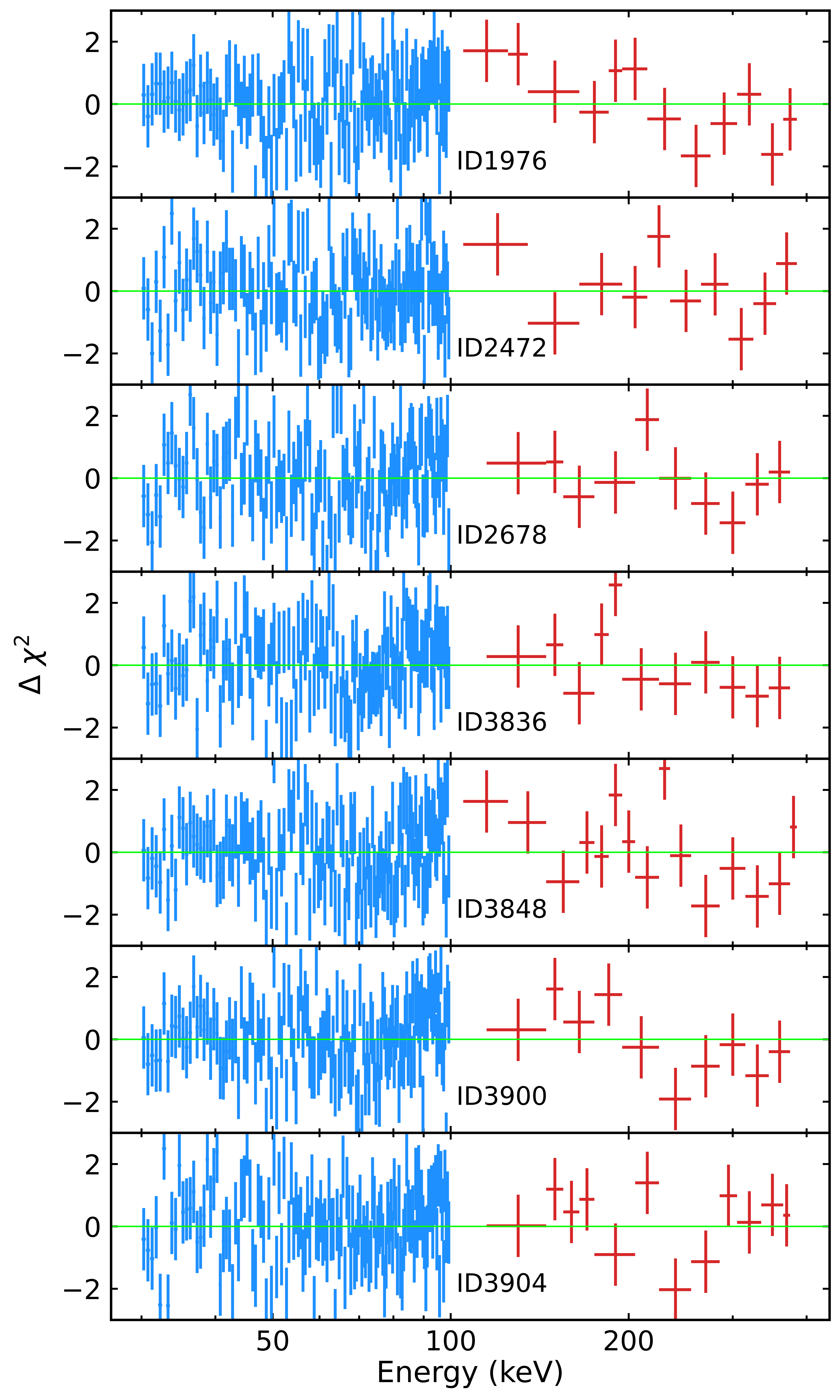}
		\caption{Residual plots of remaining observations (14). The blue and red colours are used for SE, and CS residuals, respectively. The ObsID is shown at the center of each panel.}
		\label{res_all}
	\end{figure}


\bsp	
\label{lastpage}
\end{document}